\documentclass[fleqn,usenatbib]{mnras}

\usepackage{newtxtext,newtxmath}
\usepackage{url}
\usepackage{xcolor}
\usepackage{booktabs}
\usepackage{float}
\usepackage{upgreek}
\usepackage{anyfontsize}
\DeclareMathAlphabet{\msfsl}{OT1}{cmss}{bx}{n} 

\usepackage[T1]{fontenc}
\usepackage{amsmath}

\DeclareRobustCommand{\VAN}[3]{#2}
\let\VANthebibliography\thebibliography
\def\thebibliography{\DeclareRobustCommand{\VAN}[3]{##3}\VANthebibliography}

\usepackage{graphicx}	
\usepackage{amsmath}	

\newcommand{\postreview}[1]{#1}

\newcommand{\Mth}{M_\mathrm{th}}
\newcommand{\Mp}{M_\mathrm{p}}
\newcommand{\Hp}{H_\mathrm{p}}
\newcommand{\hp}{h_\mathrm{p}}
\newcommand{\Rp}{R_\mathrm{p}}
\newcommand{\Rog}{R_\mathrm{OG1}}
\newcommand{\Rig}{R_\mathrm{IG1}}

\newcommand{\OmK}{\Omega_\mathrm{K}}

\newcommand{\Tosl}{T_\mathrm{OSL}}
\newcommand{\Fdep}{F_\mathrm{dep}}
\newcommand{\Fwave}{F_\mathrm{wave}}

\newcommand{\Sigmap}{\Sigma_\mathrm{p}}

\newcommand{\Phip}{\Phi_\mathrm{p}}
\newcommand{\FB}{\textbf{\textit{a}}_{\mathrm{B}}}
\newcommand{\FBlin}{\textbf{\textit{a}}_{\mathrm{B, lin}}}
\newcommand{\dz}{\mathrm{d}z}
\newcommand{\dphi}{\mathrm{d}\phi}
\newcommand{\dt}{\mathrm{d}t}
\newcommand{\rs}{r_\mathrm{s}}
\newcommand{\cs}{c_\mathrm{s}}
\newcommand{\bfv}{\msfsl{v}}
\newcommand{\bfu}{\boldsymbol{u}}
\newcommand{\bfa}{\textbf{\textit{a}}}

\usepackage{verbatim}

\defcitealias{cordwell_early_2024}{CR24}

\title[
How 2D are planet--disc interactions? I.]{How two-dimensional are planet--disc interactions? I. Locally isothermal discs}

\author[A. J. Cordwell et al.]{
Amelia J. Cordwell $^{1}$\thanks{E-mail: ajc356@cam.ac.uk},
Alexandros Ziampras$^{2,3,4}$,
Joshua J. Brown$^{1}$, 
and Roman R. Rafikov$^{1, 5}$
\\
$^{1}$Department of Applied Mathematics and Theoretical Physics, University of Cambridge, Wilberforce Road, Cambridge CB3 0WA, UK\\
$^{2}$ Ludwig-Maximilians-Universit{\"a}t M{\"u}nchen, Universit{\"a}ts-Sternwarte, Scheinerstr.~1, 81679 M{\"u}nchen, Germany\\
$^{3}$ Astronomy Unit, Dept. of Physics and Astronomy, Queen Mary University of London, London E1 4NS, UK\\
$^{4}$ Max Planck Institute for Astronomy, K{\"o}nigstuhl 17, 69117 Heidelberg, Germany\\
$^{5}$Institute for Advanced Study, Einstein Drive, Princeton, NJ 08540, USA
}

\date{Accepted XXX. Received YYY; in original form ZZZ}

\pubyear{2025}

\begin{document}
\label{firstpage}
\pagerange{\pageref{firstpage}--\pageref{lastpage}}
\maketitle

\begin{abstract}
Planet--disc interactions, despite being fundamentally three-dimensional, are often studied in the two-dimensional `thin-disk' approximation. The overall morphology of planet--disc interactions has been shown to be similar in both 2D and 3D simulations, however, the ability of a 2D simulation to quantitatively match 3D results depends strongly on how the potential of the planet is handled. Typically, the 2D planetary potential is smoothed out using some `smoothing length', a free parameter,  for which different values have been proposed, depending on the particular aspect of the interaction of interest. In this paper, we re-derive 2D Navier--Stokes in detail for planet--disc interactions to find better ways to represent the 2D gravitational force. We perform a large suite of 2D and 3D simulations to test these force prescriptions. We identify the parts of the interaction that are fundamentally 3D, and test how well our new force prescriptions, as well as traditional smoothed potentials, are able to match 3D simulations. 
Overall, we find that the optimal way to represent the planetary potential is the `Bessel-type potential', but that even in this case 2D simulations are unable to reproduce the correct scaling of the total torque with background gradients, and are at best able match the one-sided Lindblad torque and gap widths to level of 10 per cent. We find that analysis of observed gap structures based on standard 2D simulations may systematically underestimate planetary masses by a factor of two, and discuss the impacts of 3D effects on observations of velocity kinks. 
\end{abstract}




\section{Introduction}

\begin{figure}
    \centering
    \includegraphics[width=\linewidth]{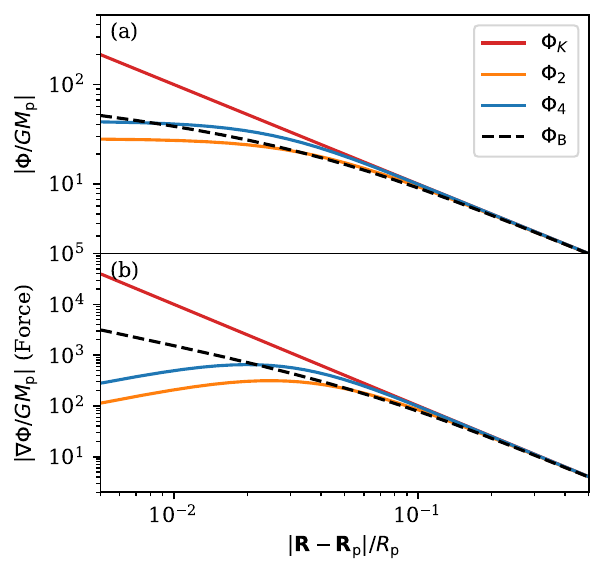}
    \caption{
    Gravitational potential (a) and acceleration (b) for the different types of planetary potential considered in this paper, shown in units $G\Mp = 1$. $\Phi_{\mathrm{K}}$ is the standard Keplerian potential, $\Phi_2$ (Equation \ref{eq:second_order_pot}) and $\Phi_4$ (Equation \ref{eq:fourth_order_pot}) are the second and fourth order smoothed potentials using a smoothing length of $r_s = bH$ with $b=0.7$. $\Phi_{\mathrm{B}}$ is the Bessel-type potential (Equation \ref{eq:besselpotl}).
    All potentials were calculated for a reference disc with a constant scale height of $H = 0.05R_p$. 
    All potentials converge to $\Phi_{\mathrm{K}}$ far from the planet, however near the planet there are significant differences, especially, in acceleration.}
    \label{fig:intro_pot}
\end{figure}

Planets have significant and complex interactions with their natal environment of protoplanetary discs including accretion, migration and gap formation \citep[for a recent review, see][]{paardekooper_planet-disk_2022}. They are one of the causes for rings and gaps seen in dust emission in protoplanetary discs \citep{andrews_observations_2020}, and the waves excited by planet--disc interactions can be observed in their kinematic signatures \citep[e.g.][]{pinte_kinematics_2022,teague_exoalma_2025}.

While the planet--disc interaction is inherently a three-dimensional (3D) process, a two-dimensional (2D) approximation is often used to simplify the complex interaction in both numerical and analytical studies. In the 2D approximation one operates with the 2D fluid equations formulated for the vertically averaged velocities in the plane of the disc (ignoring the velocity perpendicular to the disc plane) and the surface density of the disc $\Sigma$ (instead of the volumetric gas density $\rho$). Formal derivation of such 2D equations from their full 3D analogues in a disc without a planet has been provided in the past assuming the disc to be in a vertical hydrostatic equilibrium \citep{Hunter1972,Churilov1981,Goldreich1986,Ostriker1992,Fridman1996}. Such a projection of the 3D equations onto two dimensions typically results in a non-trivial modification of the pressure-density relation.

In the planet-disc context the 2D-3D correspondence has been justified in part because the majority of the angular momentum, arising from interactions with planets on non-inclined orbits, is carried in `2D waves', i.e. waves with no vertical velocity \citep{takeuchi_wave_1998}. In addition, the width of the horseshoe co-orbital region is near invariant with altitude above the midplane in vertically isothermal discs \citep{fung_3d_2015, masset_horseshoe_2016}. Moreover, gaps carved by planets are qualitatively similar in 2D and 3D simulations \citep{fung_gap_2016}.

When transitioning from 3D to 2D in the planet-disc case, one needs to worry about the vertical averaging of the gravitational force exerted by the planet on the disc. In numerical work, this averaging is often implemented approximately, by resorting to some form of the smoothed gravitational potential. Smoothing over some characteristic length $\rs$ was introduced initially \postreview{\cite[e.g.][]{takeuchi_wave_1998, kley_mass_1999, dangelo_nested_2002}} to deal with the numerical singularity of the standard Keplerian gravitational potential, $\Phi_{\mathrm{K}} \equiv - GM_p/|\mathbf{R - \Rp}|$ at $\mathbf{R} = \mathbf{R_p}$, but later its use was extended to mimic the effect of vertical averaging of the planetary gravity in 2D simulations. The specific form of the smoothed potential varies, in particular \citet{dong_density_2011} proposed several forms of the potential which differ in terms of the rate at which they converge to $\Phi_{\mathrm{K}}$ far from the planet. In numerical applications, the most often used is the second order smoothed potential,
\begin{align}
    \label{eq:second_order_pot}
    \Phi_{2} \equiv - \, \frac{G\Mp}{\sqrt{|\mathbf{R} - \mathbf{\Rp}|^2 + \rs^2}},
\end{align}
although other forms \citep[e.g. $\Phi_4$, see][and equation (\ref{eq:fourth_order_pot}) and Figure \ref{fig:intro_pot}]{cimerman_planet-driven_2021} are occasionally used as well. The smoothing length $\rs$ is often written as a fraction of the local scale height $\rs = b H$. 

The choice of $\rs$ affects various aspects of the disc-planet coupling, and its value is usually selected based on the requirement that the 2D results match a particular metric in a full 3D calculation. For example, using the standard Keplerian gravitational potential with minimal softening only to avoid numerical singularities in a 2D simulation leads to a one-sided torque excited by the planet that is twice as large as that in an equivalent 3D simulation \citep{miyoshi_gravitational_1999}. To make the value (and overall structure) of the excited torque in 2D simulations similar to 3D,  \cite{muller_treating_2012} suggested using $b = 0.7$ with the $\Phi_2$ potential, although other analyses using the same metric \citep[e.g.][]{masset_co-orbital_2002, fung_3d_2015} found slightly different optimal smoothing lengths. 

If instead of one-sided torque one uses as a metric the total (or migration) torque applied to the planet, then an optimal value of $b = 0.4$ may be more appropriate \citep{paardekooper_torque_2010, paardekooper_torque_2011}. However, \cite{tanaka_three-dimensional_2024} showed that no value of $b$ in $\Phi_2$ is capable of reproducing the scaling of the migration torque obtained in 3D with the background gradients of temperature and surface density in the disc.

Yet another metric of 2D-3D correspondence can be formulated by considering the shape of the structures forming in the disc around the planet. For example, \cite{fung_gap_2016} found that using $b = 0.5$ provides an optimal match to the shape of the gaps carved by the planet in a disc. 

Smoothed potentials different from $\Phi_2$ or $\Phi_4$ are also possible. Recently, \cite{brown_horseshoes_2024} suggested that yet another form of the potential ($\Phi_{\mathrm{B}, \Hp}$) would more effectively capture 3D physics in 2D simulations (we show the shape of their suggested potential in Figure~\ref{fig:intro_pot}). However, their analytical work was carried out in the shearing box approximation, and as such it did not include the effect of background gradients in the disc; their prescription was also not tested in simulations. 

In this paper, we re-assess the correspondence between the 2D and 3D calculations of disc-planet coupling focusing on discs with a particularly simple thermodynamics. Namely, we consider locally isothermal discs, that is discs with no vertical temperature gradient and a simple equation of state (EoS) in the form 
\begin{equation}
    P = \cs^2 \rho,
    \label{eq:iso}
\end{equation}
where $P$ is pressure and $\cs$ is the sound speed, which can be spatially varying. A specific question we aim to address is what prescription for the planetary gravitational potential ensures the closest correspondence between 2D and 3D results? 

Starting from the 3D fluid equations, we re-derive the equations of motion for an isothermal 2D disc and use these to define four different physically motivated forms of the gravitational force due to a planet. Then, we test both these forms and traditional smoothed potentials in 2D and compare results against 3D hydrodynamic simulations. In doing this we examine the impact of the force prescriptions on the torque excitation and angular momentum deposition, as well as on the structures that form in the disc.

The plan of this paper is as follows. In Section \ref{sec:integrated_ns}, we show one method to define a 2D disc from the 3D Navier--Stokes equations and then  after defining our specific disc model in Section \ref{sec:basic_equations}, describe all of the ways we represent the 2D planetary gravitational force in our simulations in Section \ref{sec:forces}. In Section \ref{sec:methods}, we describe our numerical methods and the metrics we use with which to analyse our simulations. In Section \ref{sec:results}, we present our results. We first describe which parts of the 3D simulations are unable to be represented in 2D, before comparing 2D and 3D simulations directly. In this section, we also calculate optimal smoothing lengths for traditional smoothed potentials across a range of metrics. In Section \ref{sec:discussion}, we discuss the implications of our work before summarizing our key findings in Section \ref{sec:summary}. We include additional derivations and numerical tests in the Appendix.

This paper only considers locally (and vertically) isothermal discs. \postreview{The second work in this series \citep{ziampras_in_prep} discusses} the impacts that more realistic thermodynamics have on the convergence between gap structures in 2D and 3D simulations.


\section{Reduction from 3D to 2D}
\label{sec:integrated_ns}


In this section, we perform a direct vertical averaging of the fluid equations to motivate the form of the standard 2D fluid equations in disc-planet coupling. We will use $\bfv(R,\phi,z)$ to denote 3D vector velocity, and $\bfu(R,\phi)$ to denote the 2D velocity, which we define as 
\begin{align}
    \bfu(R, \phi) &\equiv \frac{1}{\Sigma} \int^{\infty}_{-\infty} \rho \bfv ~\dz \text{,         where} 
    \label{eq:udef}\\
\Sigma(R, \phi) &\equiv \int^{\infty}_{-\infty} \rho \dz
\label{eq:Sig}
\end{align}
is the surface density. This definition of $\bfu$ is momentum-based as density weighting is adopted. We also define velocity deviation from its vertical average 
\begin{align}
    \delta \bfv(R, \phi, z) \equiv \bfv(R,\phi,z) - \bfu(R,\phi).
    \label{eq:delv}
\end{align}
Clearly, $\int_{-\infty}^{\infty} \rho \delta \bfv~\dz = 0$.

In 3D fluid equations --- Navier-Stokes and continuity --- read 
\begin{align}
& \rho\left(\frac{\partial\bfv}{\partial t}+\left(\bfv\cdot\nabla\right)\bfv\right) = -\nabla P-\rho\nabla \Phi,
\label{eq:NS}\\
& \frac{\partial\rho}{\partial t}+\nabla\cdot\left(\rho\bfv\right) = 0,
\label{eq:cont}
\end{align}
where $P$ is the 3D pressure and $\Phi$ is the full gravitational potential. Multiplying the continuity equation by $\bfv$ and adding the result to (\ref{eq:NS}), one obtains
\begin{align}
\frac{\partial (\rho\bfv)}{\partial t}+\left(\rho\bfv\cdot\nabla\right)\bfv+ \bfv \left(\nabla\cdot\left(\rho\bfv\right)\right) = -\nabla P-\rho\nabla \Phi.
\label{eq:NS1}
\end{align}
Next, we integrate this equation over $z$ from $-\infty$ to $\infty$. The first term on the left becomes $\partial (\Sigma\bfu)/\partial t$ without any additional assumptions. When manipulating the next two terms we separate spatial derivatives in the direction parallel to the disk midplane ($\nabla_\parallel$) and normal to it (i.e. $\partial/\partial z$). Then 
\begin{align}
&\int^{\infty}_{-\infty}\dz\left[
\left(\rho\bfv\cdot\nabla\right)\bfv+ \bfv \left(\nabla\cdot\left(\rho\bfv\right)\right)\right]=
\nonumber\\
&\int^{\infty}_{-\infty}\dz\left[
\left(\rho\bfv_\parallel\cdot\nabla_\parallel\right)\bfv+ \bfv \left(\nabla_\parallel\cdot\left(\rho\bfv_\parallel\right)\right)+\rho v_z\frac{\partial\bfv}{\partial z}
+\bfv \frac{\partial(\rho v_z)}{\partial z}\right].
\label{eq:terms1}
\end{align}
Last two terms combine to $\partial(\rho v_z\bfv)/\partial z$ and integrate to zero, provided that $\rho\to 0$ as $|z|\to \infty$. 

In the remaining two terns we substitute $\bfv=\bfu+\delta\bfv$:
\begin{align}
\int^{\infty}_{-\infty}\dz\Big[ &
\left(\rho\bfv_\parallel\cdot\nabla_\parallel\right)\bfu
+ \bfu \left(\nabla_\parallel\cdot\left(\rho\bfv_\parallel\right)\right)+\left(\rho\bfv_\parallel\cdot\nabla_\parallel\right)\delta\bfv 
\nonumber\\
&+ \delta\bfv \left(\nabla_\parallel\cdot\left(\rho\bfv_\parallel\right)\right)\Big]
=\left(\Sigma\bfu\cdot\nabla\right)\bfu+
\bfu \left(\nabla\cdot\left(\Sigma\bfu\right)\right)+{\bf \Delta},
\label{eq:terms2}
\end{align}
where in the first two terms (obtained without making any approximations) we dropped the $\parallel$ subscript, assuming now $\bfu=(u_R,u_\phi)$ to be the usual 2D in-plane velocity (from now on we will not be interested in $u_z$) and $\nabla=\nabla_\parallel$ to denote gradients within the disc plane. Also, along the disc plane the (vector) remainder term is
\begin{align}
{\bf \Delta} &= \int^{\infty}_{-\infty}\dz\Big[ \left(\rho\bfv_\parallel\cdot\nabla_\parallel\right)\delta\bfv_\parallel 
+ \delta\bfv_\parallel \left(\nabla_\parallel\cdot\left(\rho\bfv_\parallel\right)\right)\Big]
\label{eq:line1}\\
& =  \nabla\cdot \hat P_\mathrm{dyn},~~~~\mbox{where}~~~~\hat P_\mathrm{dyn}=\int_{-\infty}^{\infty}  \left( \rho \delta \mathbf{v_{\parallel}} (\delta \mathbf{v_{\parallel}})^{\intercal}\right) \dz.
\label{eq:Delta}
\end{align}
To obtain (\ref{eq:Delta}) we substituted $\bfv_\parallel=\bfu+\delta\bfv_\parallel$ and carried out integration over $z$ where appropriate. This term depends on the deviation of the 3D velocity along the disc plane $\bfv_\parallel=(v_R,v_\phi)$ from its vertically averaged counterpart $\bfu$. One can see that ${\bf \Delta}$ is the divergence of the 'dynamic stress tensor' $\hat P_\mathrm{dyn}$, which arises because of the vertical variation of fluid velocities in the disc.

Substituting these results into the vertically integrated equation (\ref{eq:NS1}) for the in-plane velocity components, we find 
\begin{align}
\frac{\partial (\Sigma\bfu)}{\partial t}+\left(\Sigma\bfu\cdot\nabla\right)\bfu+
\bfu \left(\nabla\cdot\left(\Sigma\bfu\right)\right)+{\bf \Delta} 
= -\nabla P_\mathrm{2D}+\Sigma~\bfa,
\label{eq:NS2}
\end{align}
where $P_{\mathrm{2D}} \equiv \int^{\infty}_{-\infty} P \dz$ is the '2D' pressure and
\begin{align}
\bfa=-\frac{1}{\Sigma}\int_{-\infty}^\infty \rho\left(\nabla_\parallel \Phi\right)~\dz
\label{eq:acc2D}
\end{align}
is the '2D gravitational acceleration'.

Next, we integrate the continuity equation (\ref{eq:cont}) over $z$. Manipulating its terms analogous to the equations (\ref{eq:terms1}), (\ref{eq:terms2}), we obtain, without any assumptions, the 2D version of the continuity equation
\begin{align}
\frac{\partial\Sigma}{\partial t}+\nabla\cdot\left(\Sigma\bfu\right) = 0,
\label{eq:cont1}
\end{align}
where again $\nabla=\nabla_\parallel$ in 2D. 

Multiplying (\ref{eq:cont1}) by $\bfu$ and subtracting it from 
(\ref{eq:NS2}), we can re-write the equation of motion (\ref{eq:NS2}) in the following form: 
\begin{align}
\Sigma\left[\frac{\partial\bfu}{\partial t}+\left(\bfu\cdot\nabla\right)\bfu\right]+{\bf \Delta} 
= -\nabla P_\mathrm{2D}+\Sigma~\bfa.
\label{eq:NS3}
\end{align}
This equation would look like the standard 2D Navier-Stokes equation if one could neglect ${\bf \Delta}$ compared to all other terms. An alternative approximate procedure for deriving the 2D disc equations via a projection of the linearized Navier--Stokes equations onto vertical modes is demonstrated in Appendix~\ref{sec:vertical_mode}.

This derivation demonstrates that the reduction of the fluid equations (\ref{eq:NS}), (\ref{eq:cont}) from 3D to the 'standard' 2D fluid equations is not perfect in general. In other words, if one obtains $\rho$ and $\bfv$ from a 3D simulation and computes $\Sigma$ and $\bfu$ from this data using definitions (\ref{eq:udef}), (\ref{eq:Sig}), the result would differ from the $\Sigma$ and $\bfu$ computed using 2D fluid equations with some assumptions regarding $P_\mathrm{2D}$ and $\bfa$. There are three reasons for that. 

First, ${\bf \Delta}$ need not be small and may provide a non-negligible contribution in the equation (\ref{eq:NS3}). Its amplitude needs to be assessed explicitly in every problem under consideration. However, in some cases, when e.g. $\bfv_\parallel$ varies with $z$ weakly near the disc midplane where $\rho$ is high (i.e. motion is 'quasi-2D'), one would expect $|\delta\bfv_\parallel|\ll |\bfv_\parallel|$ and correspondingly, ${\bf \Delta}$ being subdominant as it is quadratic in $\delta\bfv_\parallel$. Also, one expects $|{\bf \Delta}|\ll |\nabla P_\mathrm{2D}|$ when $|\delta\bfv_\parallel|\ll \cs$, which should be easy to satisfy.

Second, $P_\mathrm{2D}$ is not necessarily a simple (and similarly looking) function of $\Sigma$ as $P$ is a function of $\rho$ in 3D. If $P\propto \rho^\gamma$ with $\gamma\neq 1$, then the relation between $P_{\mathrm{2D}}$ and $\Sigma$ will involve factors depending on the vertical profile of $\rho$, which could also be time-dependent. Even when the vertical hydrostatic equilibrium is assumed, the polytropic index of $P_\mathrm{2D}(\Sigma)$ dependence has been shown to differ from $\gamma$ in both self-gravitating \citep{Hunter1972,Fridman1996} and non-self-gravitating discs \citep{Churilov1981,Goldreich1986,Ostriker1992}. However, in a disc with an isothermal EoS (Equation \ref{eq:iso}) the 2D pressure relation is also simple\footnote{\cite{brown_horseshoes_2024} found a simple relation also when considering a vertically isothermal density structure but non-isothermal pressure response to perturbations.}: $P_{\mathrm{2D}} = \cs^2 \Sigma$, even if $\cs$ is non-uniform. This is the situation we consider in this paper.

Third, in 2D one generally writes $\bfa$ as a gradient of some '2D potential', $\bfa=-\nabla\Phi_\mathrm{2D}$. Looking at equation (\ref{eq:acc2D}), this is only possible if $\rho$ and $\Sigma$ were independent of in-plane coordinates (like in a shearing sheet approximation), as in this case $\bfa$ could be written as 
\begin{align}
\bfa=-\nabla_\parallel \left( \frac{1}{\Sigma}\int_{-\infty}^\infty \rho\Phi~dz\right).
\label{eq:acc2D1}
\end{align}
However, even in this case in an evolving disc with $\rho$ changing in time, the effective 2D potential would end up being explicitly time-dependent. And in a more realistic situation where $\rho$ and $\Sigma$ vary in the disc plane, writing $\bfa$ as a gradient of some $\Phi_\mathrm{2D}$ would not even be possible. In the next section we describe several possible models for $\bfa$ is 2D, which we test later (see Section \ref{sec:results}) numerically in an attempt to optimize the 2D-3D correspondence. 


\section{Model setup}
\label{sec:basic_equations}


Next we describe the physical setup of the planet-disc problem that we focus on. We consider a 3D gaseous disc, initially in hydrostatic equilibrium, orbiting around a star of mass $M_{\star}$. The disc is described in cylindrical co-ordinates $\{R, \phi, z\}$, with the mid-plane of the disc located at $z = 0$. The disc is inviscid and non-self-gravitating. This disc is perturbed by an embedded planet, with mass $\Mp$, on a zero-inclination and circular orbit with semi-major axis $\Rp$.  

To describe the thermodynamics of the disc, we use the locally isothermal equation of state (\ref{eq:iso}).  The disc is also vertically isothermal, with its temperature depending only on $R$. In this case, prior to the introduction of a planet, the initial vertical density distribution is
\begin{equation}
\label{eq:background_rho}
    \rho_0 = \frac{\Sigma_0}{\sqrt{2 \uppi} H} \exp \left(\frac{- z^2}{2 H^2} \right),
\end{equation}
where $\Sigma_0 \equiv \int_{-\infty}^\infty \rho_0 \dz$ is the initial surface density, the scale height is $H \equiv \cs/\OmK$, $\OmK$ is the Keplerian angular frequency, and $\cs$ is the isothermal sound speed.

The initial axisymmetric surface density, temperature and sound speed profiles of the disc are
\begin{align}
\label{eq:background_state}
    \Sigma_0 = \Sigma_{\mathrm{p}} \left(\frac{R}{\Rp} \right)^{-p}, \,\,\,\,  
    T = T_{\mathrm{p}} \left(\frac{R}{\Rp} \right)^{-q}, \,\,\,\,  
    \cs = c_{\mathrm{s,p}} \left(\frac{R}{\Rp} \right)^{-q/2},
\end{align}
where the subscript `$\mathrm{p}$' indicates a value defined at $R = \Rp$, and $q$ and $p$ are constant power law indices. 

In a globally isothermal, inviscid disc with no explicit dissipation the planet-driven wake deposits angular momentum only after it shocks. For \postreview{$q \neq 0$}, the wave can gain or lose angular momentum from the background disc as it propagates through the regions of higher or lower temperature it \citep{miranda_multiple_2019}. However, for the (reasonably high) planetary mass we consider in this paper this effect is negligible in comparison to shock-driven deposition and will be ignored.


\section{Gravitational force and potential models}
\label{sec:forces}


We now list a number of models for the effective 2D planetary force or potential that we explore in this work. 

First, the explicit form (\ref{eq:background_rho}) of the initial density distribution in our disc allows us to compute a form of 2D acceleration $\bfa$ under the assumption that $\rho$ evolves away from $\rho_0$ only a little. The disc is acted on by the combined stellar and planetary gravitational potential, $\Phi = \Phi_{\star}  + \Phip$, where $\Phi_{\star} = \mathrm{G}M_{\star} /(R^2 + z^2)^{1/2}$, and 
\begin{equation}
    \Phip =  \frac{\mathrm{G}\Mp}{(s^2 + z^2)^{1/2}},
    \label{eq:pot}
\end{equation}
with $s$ being the the polar distance between a point at $(R,\phi)$ and the planet at $(\Rp,\phi_\mathrm{p})$, 
\begin{equation}
\label{eq:polar_distance}
    s \equiv \sqrt{R^2 + \Rp^2 - 2 R \Rp \cos(\phi - \phi_{\mathrm{p}})}.
\end{equation}

We can now compute $\bfa$ by vertically averaging the planetary\footnote{Formally, we should also vertically average the gravitational force due to the star. However, the effects of this averaging are only important within a few scale heights of the gravitational mass. We do not go this close to the star, and thus it is appropriate to approximate the stellar potential as Keplerian, depending only on $R$.} potential (\ref{eq:pot}) over the density distribution (\ref{eq:background_rho}). From the definition (\ref{eq:acc2D}), the gravitational acceleration due to the planet is
\begin{align}
\label{eq:fb_r}
\bfa_\mathrm{B} \cdot \hat{R} &\equiv -\frac{1}{\Sigma} \int \rho \frac{\upartial \Phi}{\upartial R} \dz \nonumber \\ 
    &= \frac{\mathrm{G} \Mp}{\sqrt{2 \uppi} H} \int  \exp\left( \frac{-z^2}{2H^2}\right) \frac{R -  \Rp \cos{(\phi - \phi_\mathrm{p})}}{\left(s^2 +z^2\right)^{3/2}} \dz \nonumber \\ 
    &= \frac{\mathrm{G}\Mp}{\sqrt{2 \uppi} H} \frac{R -  \Rp \cos{(\phi - \phi_\mathrm{p})} }{2H^2} \mathrm{e}^y\left[K_0 \left( y \right) - K_1\left( y \right) \right],
    \\
\label{eq:fb_phi}
\bfa_\mathrm{B} \cdot \hat{\phi} &\equiv -\frac{1}{R \Sigma} \int \rho \frac{\upartial \Phi}{\upartial \phi} \dz \nonumber \\ 
     &= \frac{G M_p}{R \sqrt{2 \uppi} H} \int  \exp\left( \frac{-z^2}{2H^2}\right) \frac{R \Rp \sin{(\phi - \phi_\mathrm{p})}}{\left(s^2 +z^2\right)^{3/2}} \dz \nonumber \\ 
     &= \frac{\mathrm{G}\Mp}{\sqrt{2 \uppi} H} \frac{\Rp \sin{\left(\phi - \phi_\mathrm{p} \right)} }{2H^2} \mathrm{e}^y \left[K_0 \left( y \right) - K_1\left(y \right) \right], 
\end{align}
where $y \equiv s^2/4H^2$, and $K_0$ and $K_1$ are the zeroth and first order modified Bessel functions of the second kind. This force is equivalent to that derived by \citet[see their equations 17 and 20]{muller_treating_2012},
and we will refer to it as the `Bessel-type force' throughout the text.

In Appendix~\ref{sec:vertical_mode}, we obtain a slightly different from of the gravitational acceleration $\bfa$, which follows from an alternative derivation of the 2D fluid equations, based on linear mode decomposition. To disambiguate between these two prescriptions for $\bfa$, we refer to the force derived from linear mode decomposition by the symbol $\bfa_\mathrm{B, lin}$.


\subsection{Gravitational potentials in 2D discs}
\label{sec:potentials}


The 2D gravitational force prescriptions derived above are non-conservative, i.e. cannot be described as the gradient of a potential, except in the case of a constant scale height, see equation (\ref{eq:acc2D1}). However, it may be preferable to retain the conservative nature of gravity in 2D analysis (in Section \ref{sec:results}, we will show that using the non-conservative force prescriptions leads to unexpected qualitative differences with the 3D case). In this section, we describe four different smoothed gravitational potentials that we later explore in our 2D simulations.

The first alternative potential we consider is that of \cite{brown_horseshoes_2024},
\begin{align}
    \label{eq:bessel_const_h}
    \Phi_{\mathrm{B}, \Hp} &\equiv - \frac{\mathrm{G}\Mp}{\Hp} \frac{\mathrm{e}^{y_\mathrm{p}}}{\sqrt{2 \uppi}} K_0 \left(y_\mathrm{p}\right),
\end{align}
where $y_\mathrm{p} = s^2/(4 \Hp^2)$, $\Hp = \cs(\Rp)/\OmK(\Rp)$. This version was derived in the shearing sheet approximation, and its use of $\Hp$ thus completely ignores the dependence of scale height $H$ on $R$. If $H(R)$ were indeed constant and equal to $\Hp$, then the acceleration due to this potential would be given by $\bfa_\mathrm{B}$, see equations (\ref{eq:fb_r}), (\ref{eq:fb_phi}).

Our second alternative potential is
\begin{align}
\label{eq:besselpotl}
    \Phi_{\mathrm{B}} &\equiv - \frac{\mathrm{G}\Mp}{H} \frac{\mathrm{e}^{y}}{\sqrt{2 \uppi}} K_0 \left(y\right),
\end{align}
where $y = s^2 /(4 H^2)$, as in Equation (\ref{eq:fb_phi}). This form is the same as in Equation (\ref{eq:bessel_const_h}) except we now use the scale height as a function of $R$ rather than using a fixed value of $\Hp$. This potential can be derived either by dropping higher-order coupling terms in the linear mode analysis (Appendix \ref{sec:vertical_mode}), or by simple, though not necessarily analytically supported, extension to the \cite{brown_horseshoes_2024} form.

In addition, we consider two traditional forms of the smoothed potential. We use the second order smoothed potential as in Equation (\ref{eq:second_order_pot}), and the fourth order smoothed potential \citep{dong_density_2011}
\begin{align}
    \label{eq:fourth_order_pot}
    \Phi_{4} &\equiv - \mathrm{G}\Mp \frac{s^2 + 3\rs^2/2}{\left(s^2 + \rs^2\right)^{3/2}}.
\end{align}
Both forms rely on a choice of the smoothing length, $r_s$, which we parametrize as $\rs = b \Hp$. 
The behaviour of these potentials, along with a Keplerian potential, is illustrated in Figure \ref{fig:intro_pot}.


\section{Numerical Methods}
\label{sec:methods}


We perform a suite of 2D and 3D inviscid non-linear hydrodynamic simulations using \textsc{Athena++} \citep{stone_athena_2020}. \textsc{Athena++} solves the hydrodynamic equations
\begin{align}
    \frac{\upartial \rho}{\upartial t} + \nabla \cdot (\rho \mathbf{v}) &= 0, \text{and} \\
    \frac{\upartial (\rho \mathbf{v})}{\upartial t} + \nabla \cdot (\rho \mathbf{v} \otimes \mathbf{v} + P \mathbf{I}) &= - \rho \nabla \Phi_{\star} + \rho\mathbf{g},
\end{align}
in conservative form using a Godunov scheme, where $P$ is the pressure, $\rho$ the density, and $\mathbf{v}$ is the velocity vector. $\Phi_{\star}$ is the gravitational potential of the central star. The body forces $\mathbf{g}$ are calculated using either the gradient of a planetary potential $\mathbf{g} = -\rho \nabla \Phip$ or one of the acceleration prescriptions, $\FB$ and $\FBlin$, as described in Section \ref{sec:basic_equations}.

We simulate a disc with aspect ratio $\Hp/\Rp = 0.05$ at the planetary location  and planetary mass $\Mp = 0.25 \Mth$, where $\Mth = (\Hp/\Rp)^3 M_{\star}$ is the thermal mass. The planetary mass is in the weakly non-linear regime where the theory of \cite{goodman_planetary_2001} and \cite{rafikov_nonlinear_2002} for the shocking of the planetary wake applies, however the mass is too large for linear Type I torque estimates to be valid \citep{masset_on_2006}. 
In line with previous papers, we choose not to include the indirect potential of the planet in our simulation\footnote{Recently, \cite{crida_inertial_2025} showed that the indirect potential of the planet only weakly affects Type I migration rates.}.

Our simulations span a large grid of temperature and surface density slopes.  For surface density slopes $p \in \{0, 1.5\}$ we test temperature slopes $q \in \{0, 0.5, 1.0, 1.5, 2\}$ and for $p \in \{0.5, 1, 2\}$ we run only globally isothermal, $q = 0$, simulations. 
The globally isothermal simulations were computed using \textsc{Athena++}'s isothermal equation of state.  Locally isothermal simulations were computed with the adiabatic equation of state with $\gamma = 1.0001$ and infinitely fast cooling to the background state\footnote{\cite{miranda_planetary_2019} showed that there was a significant dynamical difference between using infinitely fast cooling and using only $\gamma \rightarrow 1$ to represent local isothermality.}.

We cover the full $2\uppi$ in azimuth and use a fiducial resolution of $N_{\phi} = 3600$ (approximately 28 cells per scale height at $\Rp$) in both 2D and 3D runs. In Appendix \ref{sec:grid_convergence}, we test the numerical convergence of our simulation with respect to grid spacing. We find that while the one-sided Lindblad torque is converged at this resolution, gap spacing was not. Nevertheless, as this resolution is already finer than many other papers in the literature, and as this paper is mostly concerned with the relationship between 2D and 3D simulations, this is a fine enough resolution as long as it is the same in both 2D and 3D simulations.


To avoid spurious shocks we ramp up the planetary mass sinusoidally over 10 orbits  \citep[equivalent to the ramping used in][]{cimerman_planet-driven_2021}. We performed temporal convergence tests for a small number of our 2D simulations, and found that both torque and angular momentum deposition were converged at 10 orbits. 
In 3D, one-sided torques and horseshoe widths were converged at 10 orbits. The migration torque was not completely converged at this point, and it has been shown that multiple viscous timescales are required to reach convergence \citep[][]{dangelo_three_2010}. Therefore, we consider convergence within 5\% across a few orbits to be acceptable (when we tested a viscous simulation we found that there was much better temporal convergence in the migration torque at this point, see Appendix \ref{sec:grid_convergence} for further detail). With the exception of some unexpected structures within the horseshoe region (see Section \ref{sec:vortensity_striping}), the deposition torque density $\upartial \Fdep/\upartial R$ (see Section \ref{sec:metric_1d}) was converged within $2.5$ per cent compared to its maximum value by 10 orbits. 
At this time the maximum deviation in surface density due to gap opening was $0.3$ per cent from the background, as such there are no concerns about instabilities such as the RWI setting in.


\subsection{3D Simulations}


Our 3D simulations used spherical co-ordinates, where $\{r, \phi, \theta\}$ are the radial, azimuthal and co-latitude co-ordinates respectively. Our logarithmically spaced radial grid spans from $0.4 \Rp$ to $2\Rp$,  and has damping zones for $R < 0.45$ and $R > 1.8$. The damping zones damp density and velocity to their initial conditions over a timescale of $0.25/\OmK$. In co-latitude the grid covered $\uppi/2 - 3h_p$ to $\uppi/2$, with the planet located at $\theta = \uppi/2$ with reflecting boundary conditions. Our fiducial resolution is
$N_r \times N_{\phi} \times N_{\theta} = 900 \times 3600 \times 156$.

We tested our simulations to ensure that covering a wider region in co-latitude, either by increasing the width above the midplane of this region or by covering both the upper and lower halves of the grid, did not significantly change our results. 

Even though we do not have to physically smooth the planetary potential in 3D we still have to smooth it for practical numerical reasons. To do so we used the $\Phi_2$ potential with a smoothing length of $b = 0.1$. This choice of smoothing length has been shown to have only sub-percent effects on the horseshoe width and torques \citep{masset_horseshoe_2016}. 


\subsection{2D Simulations}


In 2D, our logarithmically spaced radial grid spans $0.2 \,\Rp$ to $4 \,\Rp$ with damping zones for $R < 0.28\,\Rp$ and $R > 3.4\,\Rp$, arranged analogous to 3D. The grid is uniformly spaced in $\phi$ across $2 \uppi$ with periodic boundary conditions. Our fiducial resolution is $N_{R} \times N_{\phi} = 1800 \times 3600$.

We ran simulations with all of the 2D potentials and gravitational accelerations that were discussed in Section \ref{sec:basic_equations}, i.e. $\FB$, $\FBlin$, $\Phi_{\mathrm{B}}$, $\Phi_{\mathrm{B}, \Hp}$, $\Phi_2$, \& $\Phi_4$.  We implemented the Bessel-type acceleration using Bessel function representation from the \textsc{cephes} Math Library \citep{moshier_methods_1989}. 
For $\Phi_2$, we tested smoothing lengths $b \in \{ 0.4, 0.6, 0.7, 0.9\}$. For $\Phi_4$, we tested smoothing lengths $b \in \{0.7, 1.1, 1.4\}$, but only performed globally isothermal simulations. A small number of additional tests were run with a wider range of smoothing lengths, however we do not include those in the discussion of this paper.


\subsection{2D-3D comparison metrics}
\label{sec:metrics}


Similarity between the 2D and 3D results can be judged based on a number of criteria. In this work we consider several comparison metrics, which can be loosely grouped into two types based on their physical nature. 

The first kind of metrics are based on the angular momentum exchange in the course of planet-disc interaction, and involves different types of torques. In locally isothermal discs, two types of torques are excited by planet--disc interactions. One are the Lindblad torques which excite density waves (a process characterized by the {\it excitation} torque density, see equations (\ref{eq:torque}) and (\ref{eq:3D_torque})) that subsequently travel radially away from the planet \citep{goldreich_excitation_1979}. These density waves carry angular momentum which they deposit into the disc when they dissipate (a process characterized by the {\it deposition} torque density, see equation (\ref{eq:f_dep_def})). In a globally isothermal or adiabatic disc, this occurs due to shock formation \citep{goodman_planetary_2001, rafikov_nonlinear_2002}, but
in non-isothermal discs, dissipation can be caused by radiation or cooling \citep{miranda_planet-disk_2020,ziampras_modeling_2023}. \postreview{(In {\it locally} isothermal discs changes in background temperature can also lead to some angular momentum dissipation 
\citep{miranda_planetary_2019}, however, for the cases considered in this paper this type of dissapation is unimportant.) }

Angular momentum exchange is also mediated by the co-rotation or horseshoe torque, which is limited in effect to a narrow region co-orbital with the planet. This torque is proportional to the local vortensity gradient, however in inviscid discs horseshoe motion quickly leads to an averaging of vortensity across this region \citep{balmforth_non-linear_2001} saturating the torque \postreview{\citep{ward_horseshoe_1991, masset_co-orbital_2002, masset_on_2006, masset_horseshoe_2016}}. In non-isothermal discs, other torques can be excited, including buoyancy torques \citep{zhu_planet-disk_2012,mcnally_low-mass_2020,ziampras_buoyancy_2024}, thermal torques due to planetary luminosity \citep{benitez-llambay_planet_2015} and radiative cooling within the horseshoe region \citep{lega_migration_2014,ziampras_migration_2024}.

The second type of metrics that we use are based on the structures that form in the disc as a result of planet-driven density wave dissipation. The dissipation of the wave angular momentum into the disc causes disc to evolve and gaps to form \citep[e.g.][hereafter \citetalias{cordwell_early_2024}]{Rafikov2002b,kanagawa_formation_2015, cordwell_early_2024}. The detailed shapes of the gaps can also be compared between the 2D and 3D simulations, providing us with another kind of metric. 

The metrics we use have different dimensionality. We use 2D and radial 1D metrics in order to understand some qualitative differences, and we also use integrated single value (summary) metrics to numerically quantify the performance of different acceleration/potential prescriptions. These summary values are also used to calculate optimal smoothing lengths for $\Phi_2$ and $\Phi_4$.

\begin{figure}
    \centering
    \includegraphics[width=\linewidth]{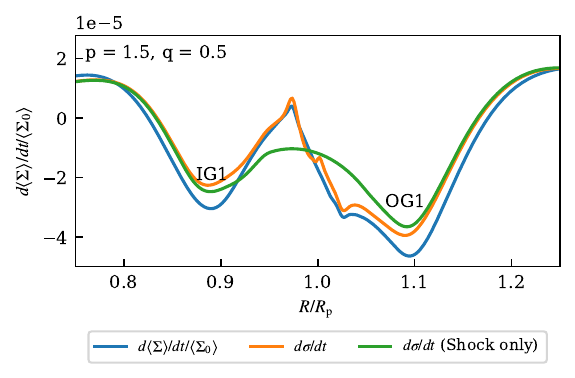}
    \caption{1D profile of the surface density evolution measured from a 3D simulation, $\langle \Sigma_0 \rangle^{-1} \, \, d \langle \Sigma \rangle/dt  $, compared to the expected $\Sigma$ evolution,  $d \sigma/dt$, using the theory of \protect \citetalias{cordwell_early_2024}, with $\Fdep$ measured from the simulation. This run used power law indices of $\Sigma$ and $T$ of $p=1.5$ and $q=0.5$. We also show the evolution expected if we set $\Fdep$ to zero within one scale height of the plane, $d \sigma/dt$ (shock only), which describes the evolution caused by the shocking of the planetary wake only (i.e. with co-rotation effects removed). The labels IG1 and OG1 refer to the locations of the inner and outer gaps closest to the planet, and are used to determine gap spacing. Despite some quantitative differences in gap depth between the measured and theoretical $\Sigma$ evolution, the theory of \protect \citetalias{cordwell_early_2024} is clearly applicable to the 3D case. }
    \label{fig:gap_definition}
\end{figure}

\begin{figure}
    \centering
    \includegraphics[width=\linewidth]{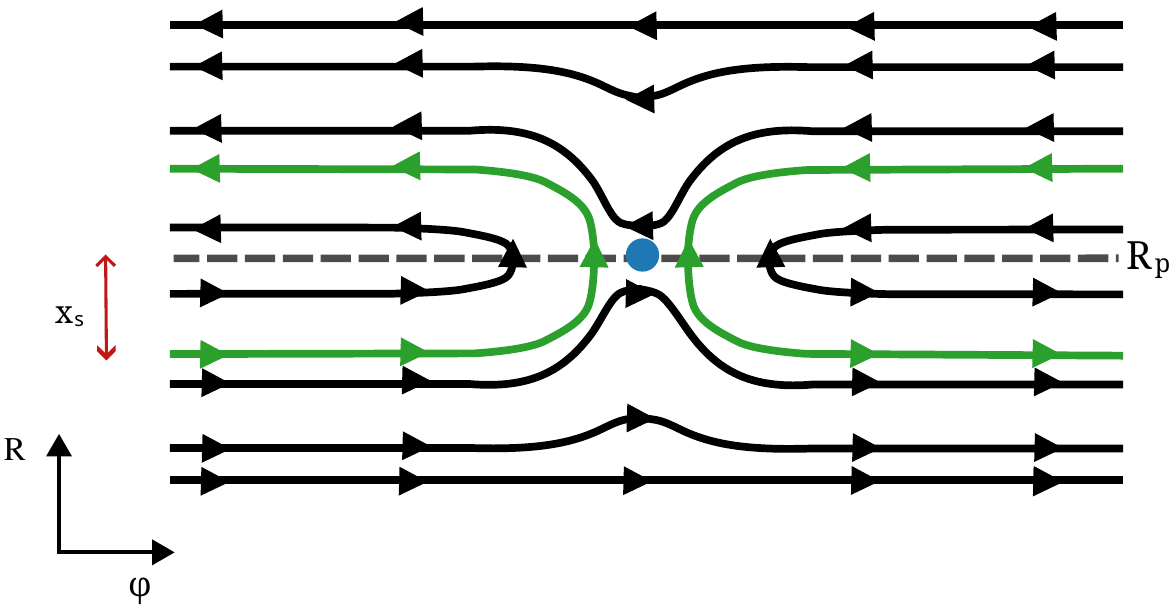}
    \caption{A sketch of fluid streamlines within the frame co-rotating with the planet. Close to the radial location of the planet the gas will undergo horseshoe turns on closed loops. The width of this region is key in determining the magnitude of the co-orbital torque \protect \cite[e.g.][]{brown_horseshoes_2024}. The horseshoe width, $x_s$, is the radial distance between $\Rp$ and the most radially distant streamline that undergoes a horseshoe turn (shown in green).}
    \label{fig:horseshoe_sketch}
\end{figure}


\subsubsection{2D Metrics}

We use two 2D metrics to understand the qualitative behaviour of the simulations. First, we use $\Sigma/\Sigma_0$, the normalised surface density, and secondly we look at vortensity, $\zeta$. In 2D simulations this is defined as $\zeta = \nabla \times \bfu /\Sigma$, and for 3D simulations we use the \cite{masset_horseshoe_2016} definition,
\begin{equation}
    \zeta_{3D} = \left[\int_{-\infty}^{\infty} \left( \frac{\rho}{ (\nabla \times \bfv)_z} \right) \dz \right]^{-1}.
    \label{eq:vort3D}
\end{equation}


\subsubsection{Radial Metrics}
\label{sec:metric_1d}

We use azimuthally-integrated 1D radial metrics to directly compare between 2D and 3D simulations:

\begin{itemize}
    \item Firstly, we consider the excitation torque density, i.e. the amount of torque excited  gravitationally by the planet per unit radius. In 2D disks this is defined as 
\begin{equation}
    \frac{\upartial T}{\upartial R} \equiv -R \oint \Sigma(R, \phi) \frac{\upartial \Phi_\mathrm{p}}{\upartial \phi} \dphi,
    \label{eq:torque}
\end{equation}
and in 3D the equivalent definition is 
\begin{equation}
    \frac{\upartial T}{\upartial R} \equiv 
     - \, R\oint\int^{\infty}_{-\infty}
    \rho \frac{\upartial \Phi}{\upartial \phi} \dz \dphi.
    \label{eq:3D_torque}
\end{equation}
For 3D discs, we also define the `2D Mode' of the torque as 
\begin{equation}
    \label{eq:torque_2d_mode}
    \frac{\upartial T}{\upartial R}\bigg|_{2D} \equiv - R \oint \left( \int^{\infty}_{-\infty} \rho \dz\right) \frac{\upartial \Phi_\mathrm{B}}{\upartial \phi} \dphi.
\end{equation}
This is derived from the lowest order mode of the vertical mode analysis in Appendix \ref{sec:vertical_mode}.
Equation (\ref{eq:torque_2d_mode}) therefore represents the amount of torque that would be excited if the vertical structure of the disc remained as in Equation (\ref{eq:background_rho}).

\item Secondly, we consider the angular momentum deposition function,
\begin{equation}
    \frac{\mathrm{d}\Fdep}{\mathrm{d}R} = \frac{\upartial T}{\upartial R} - \frac{\upartial \Fwave}{\upartial R},
    \label{eq:f_dep_def}
\end{equation}
where $\Fwave$ is the amount of angular momentum carried in the planet-driven density wave. For 2D discs, 
\begin{equation}
    \Fwave \equiv \oint \Sigma u_{R} \delta u_{\phi} \dphi,
\end{equation}
where $\delta u_{\phi} \equiv u_{\phi} - (\oint \Sigma \dphi)^{-1} \oint\Sigma u_{\phi}\dphi $
as per \citetalias{cordwell_early_2024}. In 3D discs,
\begin{equation}
    \Fwave \equiv R^2 \oint\int^{\infty}_{-\infty}  \rho v_{R}\delta v_{\phi}  \dz \dphi,
    \label{eq:f_wave_3D}
\end{equation}
where $\delta v_{\phi} \equiv v_{\phi} - \big(\oint\int^{\infty}_{-\infty} \rho \dz \dphi\big)^{-1} \oint\int^{\infty}_{-\infty} \rho  v_{\phi} \dz\dphi $.
A derivation for this metric in 3D discs is included in Appendix \ref{sec:1d_from_3d}.
\item Thirdly, we consider the azimuthally averaged surface density evolution,
\begin{equation}
    \frac{\mathrm{d} \langle \Sigma \rangle}{\dt} \frac{1}{\langle \Sigma_0 \rangle} = \frac{\langle \Sigma(t =11 P_\mathrm{p}) \rangle - \langle \Sigma(t =10 P_\mathrm{p}) \rangle }{\langle \Sigma_0 \rangle},
\end{equation}
where $\langle ...\rangle$ represents azimuthal averaging. \postreview{In the initial stages of gap-opening surface density evolution is linear in time  \citepalias{cordwell_early_2024}, and as such the period separated difference is representative of the instantaneous derivative on the LHS of this equation.}

\item Lastly, we calculate the expected surface density evolution if the co-rotation torque were saturated. The co-rotation torque saturates over the course of a few libration timescales, and as such it will not impact long term gap growth. For the planetary mass and thermodynamic equation of state used in this paper, the excited planetary wakes do not dissipate their angular momentum until they are over a scale height away from the location of the planet \citep{goodman_planetary_2001}. As such, any angular momentum deposition within a scale height of the planet will be caused by the co-rotation torques\footnote{This is a slight oversimplification. In locally isothermal discs, $\Fwave/\cs^2$, rather than $\Fwave$ as in globally isothermal or adiabatic discs, is conserved \cite{miranda_multiple_2019}. However, as we are only considering a small area of the disc, where $\cs^2$ does not change significantly, conservation of $\Fwave$ in the co-orbital region is a valid assumption.}.

\citetalias{cordwell_early_2024} derived a procedure to calculate gap evolution expected for a given function $\upartial \Fdep/\upartial R$. Therefore, we can remove the angular momentum deposition in the co-orbital region from a measured $\upartial \Fdep/\upartial R$ and use this new, 'shock-only', angular momentum deposition function to generate a `shock-only' gap evolution profile, $\mathrm{d} \sigma/\dt$ (shock only). We do this by setting $\upartial \Fdep/\upartial R$ to zero within one scale height from $\Rp$ and then passing it through the machinery of \citetalias{cordwell_early_2024}.

\citetalias{cordwell_early_2024} developed their framework for 2D discs, and only tested their theory using semi-analytic formulae for angular momentum deposition obtained in \cite{cimerman_emergence_2023}. To ensure that their machinery can be applied to 3D, we re-derive the 1D equations of angular momentum transport in 3D discs in Appendix \ref{sec:1d_from_3d}.  We also performed tests of their method using $\upartial \Fdep/\upartial R$ as measured from simulations, we show this Figure \ref{fig:gap_definition}. This figure also displays the difference between $\mathrm{d}\sigma/\dt$ (shock only) and $\langle \Sigma_0 \rangle^{-1}\mathrm{d}\langle \Sigma \rangle/\dt$.

\end{itemize}

These 1D measures are all known to be approximately constant in time for the early stages of gap opening in 2D simulations \citep{cordwell_early_2024, cimerman_planet-driven_2021}, and we tested that this temporal invariance was still true in 3D simulations.


\subsubsection{Summary Metrics}
\label{sec:summary_metrics}


Our final set of metrics are individual numbers which we use to summarize and quantify our results. The first two summary metrics are derived from the excited torque, the third relates to $\mathrm{d}\sigma/\dt$ (shock only) as defined above, and the last relates to the velocity structure in the co-rotation region. 

\begin{itemize}
    \item Our first summary metric is the total excited (migration) torque,
    \begin{equation}
        \label{eq:summary_T}
        T/T_0 = \int_0^\infty (\upartial T/\upartial R) \mathrm{d}R,
    \end{equation}
    where $\upartial T/\upartial R$ is defined in Equation (\ref{eq:torque}) or Equation (\ref{eq:3D_torque}) for 2D or 3D discs respectively. The normalization value,
    \begin{equation}
        T_0 = \left(\frac{\Mp}{M_{\star}}\right)^2 \hp^{-2} \Sigmap \Rp^4 \Omega_\mathrm{p}^2,
    \end{equation}
    is the expected scaling from linear theory \citep{goldreich_disk-satellite_1980}. A positive value of $T$ represents a torque that causes inward migration of the planet.
     For our 3D simulations we will sometimes also show the equivalent as calculated from the 2D mode, where $T_\mathrm{2D} \equiv \int_0^{\infty} (\upartial T/\upartial R)|_{2D} dR$. 
    \item The second summary metric is the one-sided Lindblad torque,
    \begin{equation}
        \label{eq:summary_tosl}
        \Tosl/F_{\mathrm{J}, 0} = \frac{1}{2} \left( \left|\int_0^{\Rp} (\upartial T/\upartial R) dR \right| + \left|\int_{\Rp}^\infty (\upartial T/\upartial R) dR\right|\right),
    \end{equation}
    where the normalization term is 
    \begin{equation}
        F_{\mathrm{J}, 0} = \left(\frac{M_p}{M_{\star}}\right)^2 \hp^{-3} \Sigma_p \Rp^4 \Omega_p^2.
    \end{equation}
    This is the classic scaling of the one-sided planet-driven Lindblad torque from \cite{goldreich_disk-satellite_1980}.
    In 3D, we also define $\Tosl|_\mathrm{2D}$ analogously to $T_\mathrm{2D}$. 
    $\Tosl$ describes the overall magnitude of the interaction and determines gap depths at viscous steady state \citep{kanagawa_formation_2015}.
    \item The third summary metric is initial gap spacing. In inviscid discs planet-opened gaps initially have a double minima structure, one on each side of the planet. The spacing between these gaps, and more specifically the dust gaps they cause, is expected to be a good probe of a planetary mass from observations \citep{dong__observational_2015}. For this metric we adopt
    \begin{equation}
        \frac{\Rog - \Rig}{\Rp},
    \end{equation}
    where $\Rog$ is the closest local minima to the planet in $\sigma$ for $R > \Rp$, and $\Rig$ is the equivalent for $R < \Rp$. We show an example of gap structures, with these locations labeled, in Figure \ref{fig:gap_definition}.
    As gap opening is a secular long term effect, the short term saturated horseshoe torque will not strongly affect the overall gap structure. Therefore we calculate gap spacing from $d \sigma/dt$ (shock only) so that it represents only Lindblad wake driven evolution.
    
    \item Lastly, we consider the horseshoe half-width, $x_s$ \citep[e.g.][]{masset_horseshoe_2016}. In the frame co-rotating with the planet there exist a class of streamlines near $\Rp$ that, instead of passing the location of the planet azimuthally, perform a horseshoe turn and reverse their azimuthal velocity. We show a schematic of these streamlines in Figure \ref{fig:horseshoe_sketch}. The horseshoe half-width is the radial distance between $\Rp$ and the furthest streamline that undergoes one of these horseshoe turns.
    We calculate this by evaluating streamlines from (vertically averaged) velocities calculated using SciPy's solve\_ivp function.
\end{itemize}


\subsection{Analysis software pipeline}
\label{sec:software}


In order to calculate all of the above described metrics, we developed an interoperable analysis software pipeline that works for 2D and 3D simulations. So far, this has been tested on both \textsc{Athena++} and \textsc{PLUTO} simulations. The package is easily extendable to include output from other simulation codes, such as \textsc{FARGO3D}, as well as the inclusion of other physics (e.g. moving or eccentric planets). To allow other researchers to easily use the same metrics we have made this package available on GitHub, \url{https://github.com/cordwella/disc_planet_analysis}, under the MIT Open Source License. 


\section{Results}
\label{sec:results}


\begin{figure}
    \centering
    \includegraphics[width=\linewidth]{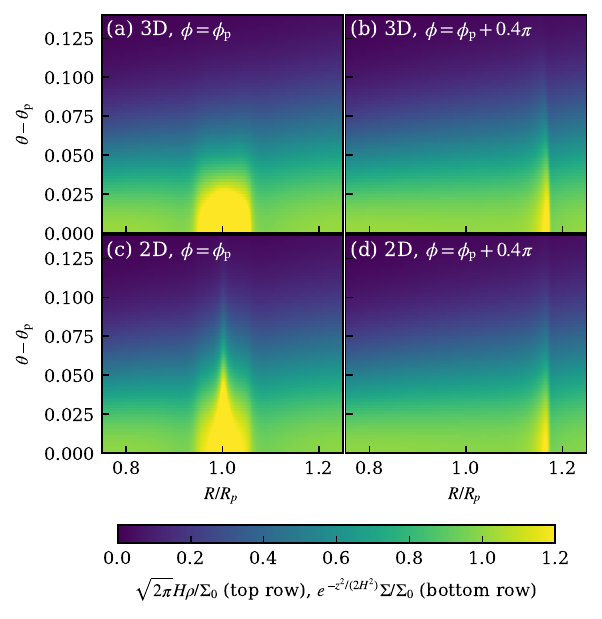}
    \caption{
    Vertical density slices from a $p=1.5$ and $q=0$ 3D simulation along lines of constant azimuth. Panels (a) \& (b) show the full 3D structure, which is compared with the density structure derived from an equivalent 2D model in panels (c) \& (d). The structure derived from 2D model is calculated as the surface density of the 3D model multiplied by $\mathrm{e}^{-z^2/2H^2}$, i.e. the initial hydrostatic vertical density (\ref{eq:background_rho}). In (a) and (c), where hydrostatic envelope around the planet is shown, it is clear that the 2D  model misrepresents the planetary atmosphere. In (b) and (d), we compare the structure of the planet induced wake at a distance away from the planet. In the full 3D case, the wave curves slightly backwards towards the planet at higher altitudes whereas in the quasi-2D case the wake is required to be vertically straight.}
    \label{fig:r-z_plane}
\end{figure}

\begin{figure}
    \centering
    \includegraphics[width=\linewidth]{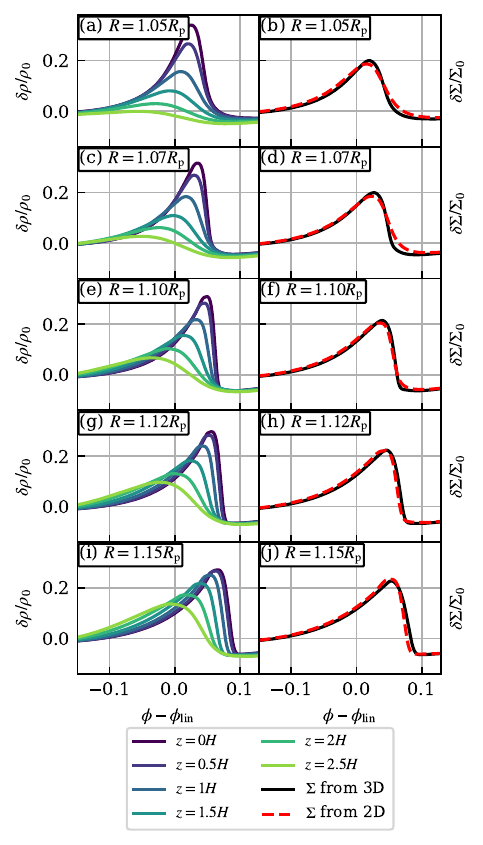}
    \caption{Left: Planet induced wake structure as a function of altitude above the midplane in the outer disc as measured from a 3D simulation ($p =1.5, q=0$). Right: Wake structure in surface density for the same 3D simulation compared to that from a 2D simulation using the potential $\Phi_2, b=0.7$. The 3D wake structure is dependent on altitude, and shocks, identified visually as when the gradient of $\delta \rho/\rho_0$ reaches its maximum,
    form at the midplane radially sooner (e) than at higher altitudes.
    The exact formation of the shock is not captured by the surface density evolution, and the literature standard 2D simulation shows slightly different shock formation and structure compared to the 3D case. 
    }
    \label{fig:shock_height}
\end{figure}

\begin{figure}
    \centering
    \includegraphics[width=\linewidth]{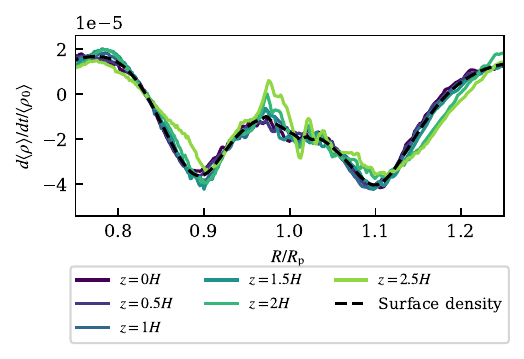}
    \caption{Time derivative of density from our fiducial 3D simulation as a function of altitude above the midplane (solid lines) compared to the time derivative of surface density (dashed line). Within a scale height of the midplane, density evolution is very similar to surface density evolution however at higher altitudes gap formation in the outer disc happens further away from the planet. 
    }
    \label{fig:ev_height}
\end{figure}

This section, which describes our simulation results, is organised as follows. First, we describe the qualitative features of 3D simulations and show how these differ from what is represented by either the vertical average of that 3D simulation or a ``literature standard'' 2D simulation that uses the second order potential $\Phi_2, b= 0.7$. 
Second, we show results from 2D simulations using the new physically motivated 2D gravitational acceleration prescriptions from Section \ref{sec:integrated_ns}, and compare these to 3D simulations. Lastly, we show results from 2D simulations using existing smoothed potentials, $\Phi_2$ and $\Phi_4$, and use these to calculate optimal smoothing lengths for different physical features. We show results measured at 10 planetary orbits from the start of the simulation. In Appendix \ref{sec:grid_convergence}, we check that this time is sufficient for temporal convergence. 


\subsection{The 3D nature of planet--disc interactions}
\label{sec:results_3D}


\begin{figure}
    \centering
    \includegraphics[width = \linewidth]{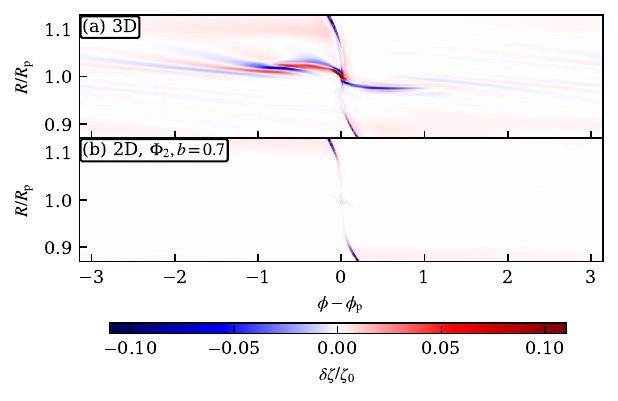}
    \caption{Map of vortensity $\zeta_\mathrm{3D}$ defined defined by equation (\ref{eq:vort3D}) based on our fiducial 3D simulation (a) compared with the vortensity map from an equivalent 2D simulation (b). In 2D, vortensity $\zeta$ is constant within the horseshoe region. However in the 3D simulation, stripes of vortensity are generated in the horseshoe region near to the planet.}
    \label{fig:3d_vortensity}
\end{figure}

To situate ourselves in the problem, we first examine the key features of the 3D planet-disc interaction that are or are not reproduced by a 2D calculation..

In Figure \ref{fig:r-z_plane}, we show azimuthal slices of the density distribution from a 3D simulation (top row) and compare this to the 'reconstructed' density structure (bottom row) that would be implied by a 2D calculation if one were to make an assumption of a vertical hydrostatic equilibrium. In practice, this means that one assumes $\rho(z)=\rho_0(z)$ given by the equation (\ref{eq:background_rho}) with $\Sigma_0$ equal to the local $\Sigma$ from a 2D run. Such 3D density 'reconstruction' from 2D data has been attempted in the past \citep[e.g][]{Zhu2015,Dong2017}.

Two differences between the true and reconstructed $\rho$ should be emphasized. First, the reconstructed model is unable to account for the near-spherical planetary atmosphere that forms in 3D, see panels (a) and (c). Second, the form of the density wave in panel (b) curves towards the planet as altitude increases \postreview{(see also Figure \ref{fig:shock_height})} --- obviously this is not reproduced in 2D (panel (d)), and therefore causes the 2D equivalent to form a less distinct shock edge.

In Figure \ref{fig:shock_height} we investigate the structure of the planet-driven density wake in more detail. In the 3D case, the relative amplitude of the density perturbation $\delta\rho/\rho_0(z)$ strongly decays with the height $z$ above the midplane, at all $R$. This is caused by weakening of the planetary potential, to which the wake couples at excitation, with $z$. As the distance from the planet increases, the wake steepens due to nonlinear effects at all heights, but much faster closer to the midplane. As a result, when the wave eventually shocks\footnote{
In this paper we visually identify a shock location by when the gradient of $\delta \rho/\rho_0$ reaches its maximum and there is a clear 'knee' break between the front of the wake and the background disc. In a 2D isothermal disc it is possible to use the perturbed vortensity structure to identify the location that the shock forms \citep[][]{cimerman_planet-driven_2021}, however this is infeasible to use in 3D due to the formation of `horseshoe vortensity stripes', see Section \ref{sec:vortensity_striping}.}, 
this happens first near the midplane, where the wave is more nonlinear. The wake eventually shocks at all measured altitudes, however the distance away from the planet at which the shock occurs increases with altitude. 

In the right column we show how the wake shows up in $\Sigma$ and compare it to a 2D simulation. Despite the wake from both types of simulations being similar in magnitude one scale height away from the planet (Figure \ref{fig:shock_height} (b)), in the 3D simulation the wake shocks earlier, (Figure \ref{fig:shock_height} (d)), than in the 2D simulation (Figure \ref{fig:shock_height} (h)).

To ensure that 3D gap evolution occurs similarly to 2D gap formation, we illustrate density evolution, expressed as a time derivative of $\rho(z)$, as a function of altitude $z$ in Figure \ref{fig:ev_height}. We find mostly very similar evolution at different altitudes, although the outer gap forms further away from the planet at higher altitudes. This difference is likely due to shock formation occurring at further distances from the planet at higher altitudes, as shown in Figure \ref{fig:shock_height}, and therefore angular momentum deposition occurring further away from the planet. There are also unexpected features, more present at higher altitudes, which we believe to be caused by 3D motion, and in particular what we will call `horseshoe vortensity striping', discussed next.


\subsubsection{Horseshoe vortensity striping}
\label{sec:vortensity_striping}

In Figure \ref{fig:3d_vortensity}, we show a maps of vortensities $\zeta$ and $\zeta_\mathrm{3D}$ (defined by equation (\ref{eq:vort3D})) from a pair of 2D and 3D simulations. Both simulations have an initially flat vortensity ($p=1.5$) and temperature profile. In 2D the bulk vortensity only evolves after the planetary wake has shocked (i.e. $R < 0.9 \Rp$), as expected from previous work \citep[e.g.][]{Li2005,Dong2011,cimerman_planet-driven_2021}. However in 3D, large scale stripes of 3D vortensity $\zeta_\mathrm{3D}$ form in the horseshoe region, see panel (a). We verified that we see such stripes also in the map of $z$-component of the conventional vortensity $\zeta_z$ at $z=0$. These features are especially prominent in the region where fluid parcels exit their horseshoe turn close to the planet (in this simulation the horseshoe region ranges from $0.975\Rp$ to $1.025\Rp$). We allowed this simulation to run for longer and found that while the shape of the structure did not change in time, it grew in magnitude. In surface density, this feature can also be seen in non-axisymmetric stripes of material, but the magnitude of this effect is much smaller than that of shock driven gap opening. 

Horseshoe vortensity striping has been previously observed by \cite{masset_horseshoe_2016} and \cite{chametla_planetary_2025}. They claimed that this feature is caused by horseshoe streamlines bending vertically down to the planet at closest approach. Vortensity is a conserved quantity in 2D, a property that would hold if planet--disc interactions remained fundamentally 2D in 3D simulations. Clearly this interaction is 3D enough for vortensity not to be conserved, which may have implications for features that have long been studied from the perspective of vortensity, such as the dynamical co-rotation torque \citep{masset_runaway_2003}.

As horseshoe vortensity striping has not yet been widely discussed, we also undertook a series of tests to ensure that it was not numerical in origin.  The structure was unchanged by differences in planetary ramp up time, or the nature and position of vertical boundaries. Using a smaller planetary mass changes the number of stripes but does not stop the effect from occurring. Including a viscosity in the simulation (we tested $\nu = 2.5 \times10^{-6}$), increasing the 3D smoothing length, or disconcertingly, increasing the local grid resolution, all caused the magnitude of this structure to decrease. This feature is clearly not numerical, although its convergence properties remain to be established.

\begin{figure}
    \centering
    \includegraphics[width=\linewidth]{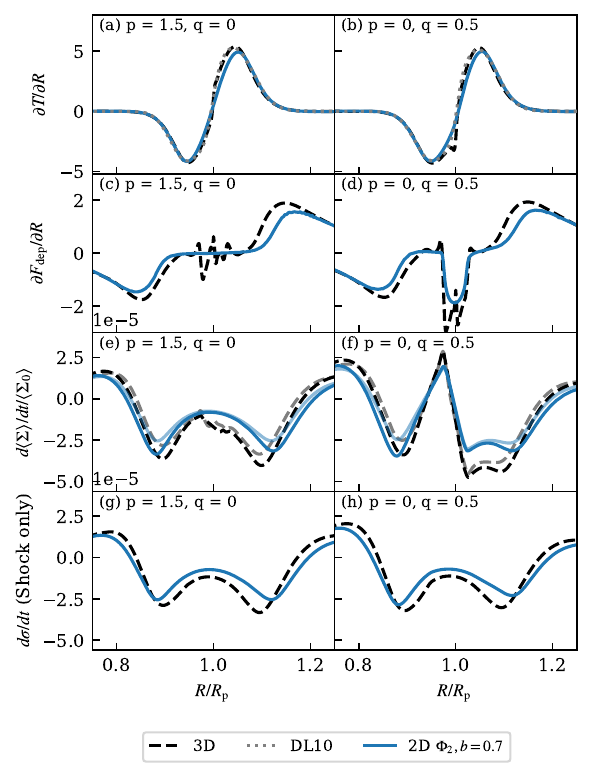}    
    \caption{
    Radial metrics as measured from 3D (dashed lines) and 2D (solid lines) simulations. The 2D simulations use second order potential $\Phi_2, b=0.7$. The simulations in the left-hand column have background surface density and temperature gradients of $p=1.5$ and $q=0$. Those in the right-hand column have $p=0$ and $q=0.5$.
    Each row shows a different radial metric as defined in Section \ref{sec:metric_1d}. The first two rows show the torque excited by the planet and angular momentum deposition  respectively (in units of $F_{\mathrm{J}, 0}$). The first row also shows the expected excitation profile from the viscous simulations of \protect \cite{dangelo_three_2010}. 
    The third row shows surface density evolution as measured from the simulation, and in matching lighter colours the expected evolution using the theory of \protect \citetalias{cordwell_early_2024} with the measured $\Fdep$ from the row above.  
    The final row shows the expected surface density evolution if 
    $\upartial \Fdep/\upartial R$ was zero within $1H$ of the planet.
    See Section \ref{sec:metric_1d} for further detail on the metrics and the text for further discussion.
    }
    \label{fig:3D_1D_torque_summary}
\end{figure}


\subsection{Radial metrics in 3D discs}


Having discussed some fundamentally 3D parts of the planet--disc interaction, we now shift our focus to considering the radial metrics as defined in Section \ref{sec:metric_1d}.
In Figure \ref{fig:3D_1D_torque_summary}, we compare representative 3D simulations with 2D simulations that use the second order potential $\Phi_2$ with $b =0.7$. The expected difference between the two representative choices of background surface density and temperature profiles is whether or not the co-rotation torque should be active. Indeed,  for the $p = 1.5, q=0$ simulations the co-rotation torque is not active and angular momentum deposition peaks away from the planet, at $R \approx 0.85$ and $R \approx 1.1$ in Figure \ref{fig:3D_1D_torque_summary}(c), as expected from weakly non-linear shocking \citep{goodman_planetary_2001, rafikov_nonlinear_2002, cimerman_planet-driven_2021}. Whereas in the $p=0, q=0.5$ simulations the co-rotation torque is active and unsaturated, leading to angular momentum deposition in the horseshoe region. However, once the effect of the co-rotation torque is removed as in panels (g), (h), the expected gap profiles are similar for both simulations regardless of their $p$ and $q$.

When comparing 2D and 3D simulations, the only qualitative difference between is due to horseshoe vortensity striping, the effect of which is limited to near the co-rotation region. At the quantitative level, there are differences dependent on the choice of smoothed potential, which we will investigate in Sections \ref{sec:two_d_bessel} and \ref{sec:2d_trad_smoothing}. However the analysis of the fundamental 3D structure of the planet-generated wake and shock suggest that even 2D simulations that correctly reproduce the initial excited structure of a planetary wake are unable to quantitively capture the correct shock evolution, and therefore angular momentum deposition profile. Generally this leads to angular momentum deposition closer to the planet than in the 2D case thus leading to slightly smaller spacings between gaps.

We also compare our torque excitation with the models of \cite{dangelo_three_2010}. In Figure \ref{fig:3D_1D_torque_summary}(b), while our 3D $\upartial T/\upartial R$ and the reference model of \cite{dangelo_three_2010} match well for the majority of the radial domain, our simulation shows an additional spike in excited torque for $R$ just under $\Rp$. On further investigation we saw this feature in the low viscosity simulations of  \cite{dangelo_three_2010} (Fig. 5), and given that this is a 3D-only feature we believe it may be related to the horseshoe vortensity striping as described above. 

We also checked how much angular momentum flux was carried in 2D versus 3D wave structures, i.e. the wave as measured from the vertically averaged velocity and density structures compared to the total wave angular momentum flux. We found that the 2D wave carried around 90\% of the total wave angular momentum flux \citep[matching theoretical predictions from][]{takeuchi_wave_1998}, but that this proportion was not radially constant. Angular momentum can be transferred between different vertical modes of the wave, and so the 2D wave cannot be treated as independent of higher order vertical modes, i.e. changes in vertical structure.

In Appendix \ref{sec:slope_on_gap}, we use our large grid of 3D simulations to show how all radial metrics are affected by changing the slope parameters $p$ and $q$.


\subsection{2D simulations with Bessel-type forces and potentials}
\label{sec:two_d_bessel}


\begin{figure}
    \centering
    \includegraphics[width=\linewidth]{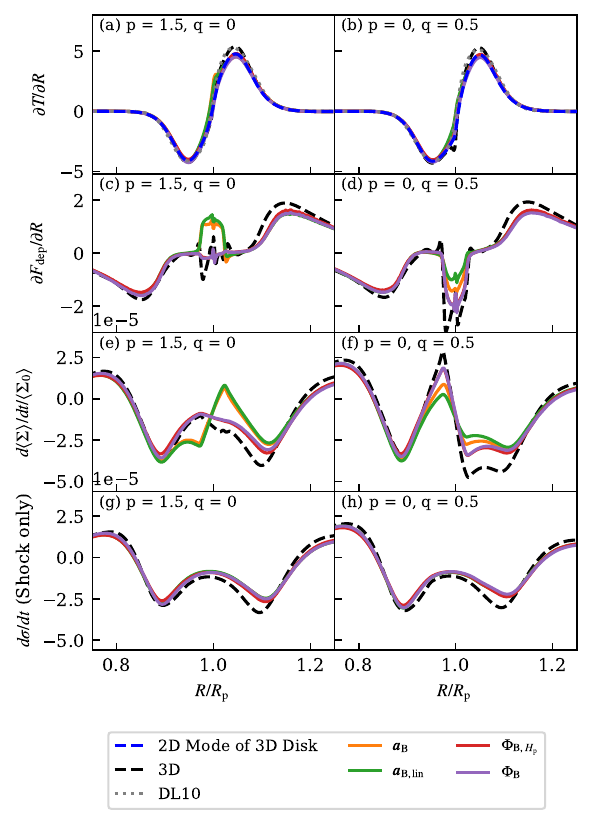}
    \caption{Same as Figure \protect \ref{fig:3D_1D_torque_summary}, except comparing the  3D results with 2D results using different Bessel-type forces (as defined in Section \ref{sec:integrated_ns}). In panels (a) and (b) we also include the 2D mode of the torque from the 3D simulations as defined in Equation (\ref{eq:torque_2d_mode}). From top to bottom the rows represent excited torque, angular momentum deposition, measured surface density evolution, and expected surface density evolution in the absence of horseshoe effects.}
    \label{fig:results_1D}
\end{figure}

\begin{figure}
    \centering
    \includegraphics[width=\linewidth]{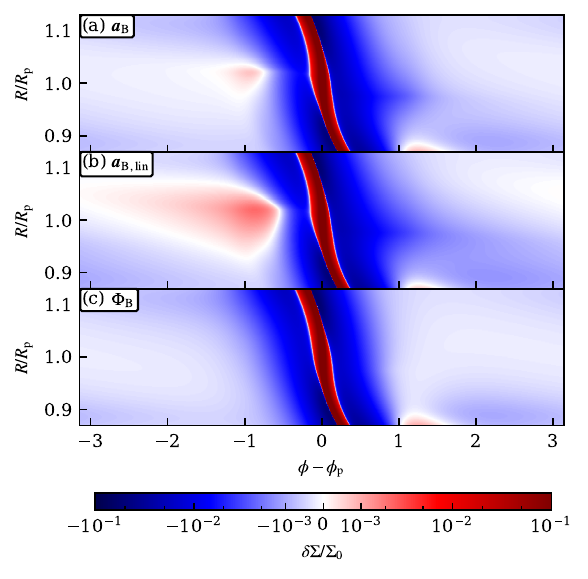}
    \caption{
    Maps of surface density in the co-orbital region around the planet for three different 2D simulations, with background gradients  $p = 1.5, q = 0$, using gravitational force descriptions of $\FB$, $\FBlin$ and $\Phi_{\mathrm{B}}$ respectively. 
    For this background state the co-rotation torque is expected to not be excited, and all of the evolution (once the density wakes are set up) should be driven by the shocks that occur outside this area. These discrepancies suggest that $\FB$ and $\FBlin$ are inappropriate choices for future 2D simulations, whereas $\Phi_{\mathrm{B}}$ is a qualitatively appropriate choice.}
    \label{fig:results_2D}
\end{figure}

\begin{figure}
    \centering
    \includegraphics[width=\linewidth]{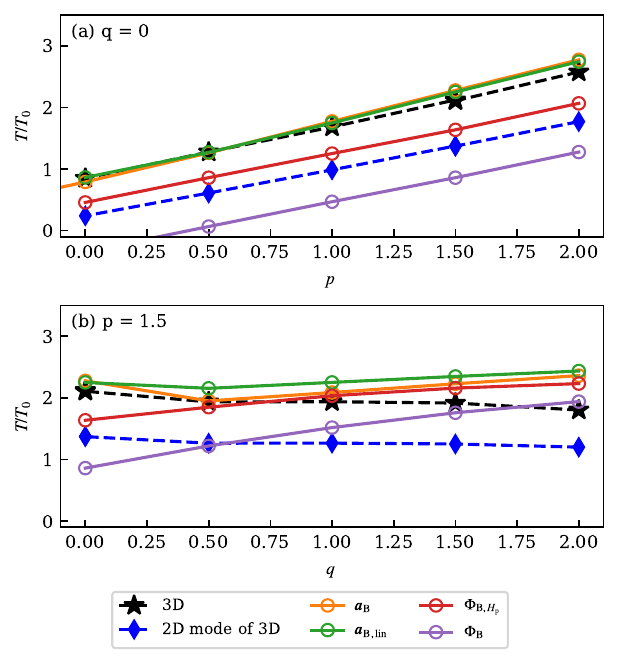}
    \caption{Total (migration) torque on the planet from 2D simulations using different Bessel-type forces (solid lines) compared to that from 3D simulations (dashed lines) as a function of surface density gradient $p$ for a globally isothermal disc (a) and of temperature gradient $q$ (b). The black dashed line and stars represent the 3D torque as measured directly from the simulation whereas the blue line and diamond represent the 2D mode of the torque (see Section \ref{sec:summary_metrics} for a description). $\FB$ is able to give the best 2D representation of the 3D migration torque, however, as shown in Figure \protect \ref{fig:results_2D}, it is an inappropriate choice to match the qualitative features of surface density evolution.}
    \label{fig:total_torque_bessel}
\end{figure}

\begin{table}
\caption{
Comparison of key values from 2D simulations using the Bessel-type gravitational forces with 3D simulations. The values presented here have been averaged across simulations with different background gradients of $p$ and $q$. The migration torque is omitted as it has a strong dependence on background gradients. Brackets indicate the torque from the 2D mode of the 3D simulation. 
}
\label{tab:bessel_av}
\begin{tabular}{lrrr}
\toprule
 & $\Tosl/F_{\mathrm{J}, 0}$ ($\Tosl|_\mathrm{2D}/F_{\mathrm{J}, 0}$) & $(R_\mathrm{OG1} - R_\mathrm{IG1})/R_\mathrm{p}$ & $x_\mathrm{s}$ \\
\midrule
3D & 0.379 (0.345)& 0.191 & 0.025 \\
$\FBlin$ & 0.338 & 0.215 & 0.024 \\
$\FB$ & 0.338 & 0.215 & 0.024 \\
$\Phi_{\mathrm{B}}$ & 0.339 & 0.215 & 0.024 \\
$\Phi_{\mathrm{B}, \Hp}$ & 0.341 & 0.215 & 0.024 \\\bottomrule
\end{tabular}
\end{table}

We now consider how well 2D simulations, using physically motivated (Bessel-type) gravitational accelerations, compare with 3D simulation results. In Section \ref{sec:integrated_ns}, we described two physical forms of the 2D gravitational force due to a planet from the 3D Navier--Stokes equations, $\FB$ and $\FBlin$, and two gravitational potential functions inspired by those, $\Phi_\mathrm{B}$ and $\Phi_{\mathrm{B}, \Hp}$. Our naive computational implementation for the Bessel-type forces and potentials leads to around a $20 $ per cent computational overhead compared to the second order smoothed potential. 

We display radial metrics for a selection of these simulations in Figure \ref{fig:results_1D}. Panels (a) and (b) show a good match between 2D and 3D simulations in the shape of $\upartial T/\upartial R$, although generally the 2D simulations have a lower peak torque excitation in the outer disk. When we consider the 2D mode of the 3D torque, $\upartial T/\upartial R|_{\mathrm{2D}}$ given by equation (\ref{eq:torque_2d_mode}), we find that these torque structures have an even better match. The small, but not inconsequential difference, is expected and represents the fact that gravitational acceleration and excited torque are affected by perturbations in the vertical structure of the disc, see Section \ref{sec:integrated_ns}. We should, therefore, not expect 2D simulations, which assume no perturbation to the disc vertical structure, to be able to completely match torque values from 3D discs.

In angular momentum deposition, Figure \ref{fig:results_1D} (c, d), simulations using $\FB$ and $\FBlin$ strongly deviate from behaviour expected in the co-rotation region from 3D runs. No 2D simulation is able to represent the effect of horseshoe vortensity striping, however using $\bfa_\mathrm{B}$ and $\bfa_\mathrm{B,lin}$ results in yet additional density substructures in the co-orbital region, as we illustrate further in Figure \ref{fig:results_2D}. With the exception of the Lindblad wake centred at $\Rp$, the surface density in the co-rotation region should be axisymmetric, but both $\FB$ and $\FBlin$ cause the formation of over and under-densities which seem to be centred around the location of the widest horseshoe turn. It is likely that these occur due to the non-conservative nature of the 2D forces used -- in 3D these additional force components may be what leads to the horseshoe vortensity striping feature. We allowed some of these simulations to run for longer to ensure that this non-axisymmetric feature wasn't a transient effect. While the non-axisymmetric lobes are less visible in 2D plots at later times, the effect on angular momentum deposition and the azimuthally averaged surface density evolution was unchanged. Due to this effect, we conclude that neither $\FB$ nor $\FBlin$ should be used to represent planet--disc physics in a 2D simulation.

On the other hand, it is worth noting that when we re-calculate expected surface density evolution with the co-rotation effects removed, Figure \ref{fig:3D_1D_torque_summary}(g, h), we see that the gap profiles are very similar for all Bessel-type accelerations and potentials.

2D simulations using $\Phi_{\mathrm{B}}$ and $\Phi_{\mathrm{B, \Hp}}$ have, with the exception of missing the vortensity striping feature, good qualitative agreement with 3D simulations for all the metrics shown in Figure \ref{fig:results_1D}. Quantitively, they miss some torque excitation, which must be associated with 3D vertical modes, and their predicted gap locations are slightly further away from the planet, but insomuch as possible for a 2D simulation, they represent the physics of the interaction well.

Out of our summary metrics we found that $\Tosl$, gap spacing, and horseshoe width, $x_s$, were mostly invariant under the choice of surface density and temperature slope, whereas -- as expected from previous work and linear theory -- the migration torque is strongly dependent on $p$ and $q$. We show the average summary metrics in Table \ref{tab:bessel_av}, with the exception of migration torque. All of our Bessel-type forces underestimate the one-sided Lindblad torque by around $10$ per cent, although if we compare them instead to the 2D mode of the 3D torque they match within $3$ per cent. Gap spacings from 2D simulations tend to be overestimated by around $8$ per cent compared to 3D simulations, and 2D horseshoe widths are underestimated by $8$ per cent.

In Figure \ref{fig:total_torque_bessel}, we compare migration torques between 3D and Bessel-type 2D simulations. $\FB$ and $\FBlin$ provide the best matches to the 3D migration torques, however these misrepresent much of the other physics of the interaction. Neither $\Phi_{\mathrm{B}}$ nor $\Phi_{\mathrm{B}, \Hp}$ matched migration torque values well, although they were able to match the gradient of the torque with respect to $p$.  None of these 2D prescriptions were able to match the gradient of the torque with respect to the background temperature slope $q$.


\subsection{2D results using traditional smoothed potentials}
\label{sec:2d_trad_smoothing}


\begin{figure}
    \centering
    \includegraphics[width=\linewidth]{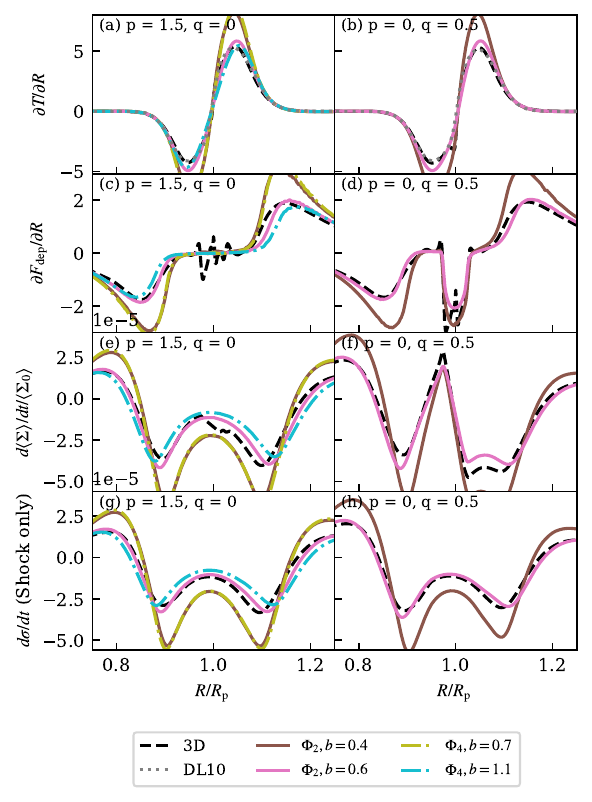}
    \caption{Same as Figure \protect \ref{fig:3D_1D_torque_summary}, except these now compare the same 3D results with 2D simulations using the second or fourth order smoothed potentials with different smoothing lengths. The fourth order potential was only used in globally isothermal simulations and as such is not included in the right column. As before, from top to bottom the rows represent excited torque, angular momentum deposition, measured surface density evolution, and expected surface density evolution in the absence of horseshoe effects.}
    \label{fig:results_1D_smoothed}
\end{figure}

\begin{figure}
    \centering
    \includegraphics[width=\linewidth]{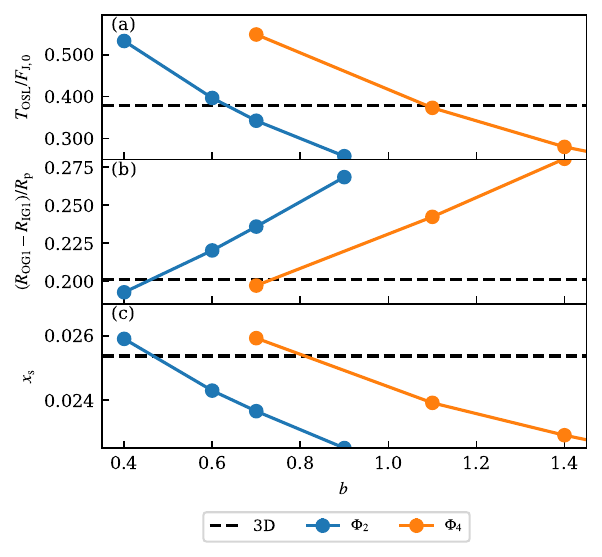}
    \caption{
    Summary metrics from 2D simulations using 
    second, $\Phi_2$, or fourth, $\Phi_4$, order potentials as a function of normalised smoothing length, $b = r_{\mathrm{s}}/H$ (solid lines), compared to results from 3D simulations (dashed black line). 
    As none of these three metrics had a systemic dependence on $p$ or $q$ the values presented here have been averaged across simulations with different background gradients $p$ and $q$. }
    \label{fig:val_over_b_second}
\end{figure}

\begin{figure}
    \centering
    \includegraphics[width=\linewidth]{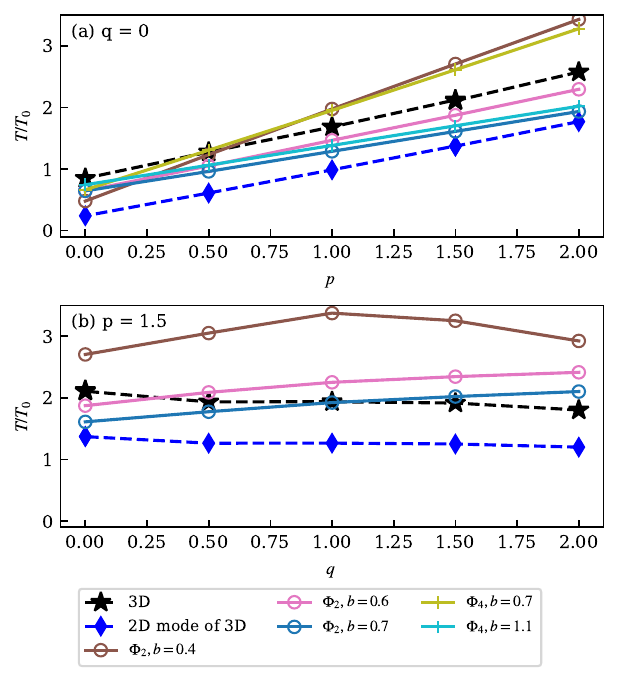}
    \caption{
    Migration torques as in Figure \protect \ref{fig:total_torque_bessel}, but with 2D simulations that use the more traditional second, $\Phi_2$, or fourth, $\Phi_4$, order smoothed potentials compared with 3D simulations. The fourth order potential, $\Phi_4$, was only used in globally isothermal simulations and as such has not been included in the lower panel.}
    \label{fig:torque_over_b_second}
\end{figure}

\begin{table}
    \label{tab:optimal_b}
\caption{Optimal smoothing lengths, $b = r_{\mathrm{s}}/H$, for traditional potentials in 2D simulations, calculated by comparison to the same metric in 3D simulations. No optimum $b$ is provided for migration torque as the optimum is strongly dependent on the choice of background gradients.}
\begin{tabular}{lccccc}
         & \multicolumn{1}{l}{$T$} & \multicolumn{1}{l}{$\Tosl$} & \multicolumn{1}{l}{$\Tosl$ (2D Mode)} & \multicolumn{1}{l}{$(\Rog - \Rig)/\Rp$} & \multicolumn{1}{l}{$x_\mathrm{s}$} \\
         \midrule
$\Phi_2$ & -                                & $0.64$                                         & $0.70$                                                     & $0.49$                            &  $0.50$                                   \\
$\Phi_4$ &     -                             &                 $1.09$                             &                       $1.19$                                 &       $0.73$                           &                                    $0.84$
\end{tabular}
\end{table}

In this subsection, we consider how well the second, $\Phi_2$, and fourth order, $\Phi_4$, smoothed potentials, using a range of smoothing lengths, are able to match the physics of 3D simulations. \cite{muller_treating_2012} performed this analysis in a form restricted to the shape of torque excitation and the second order potential only. In addition to showing $\Phi_2, b = 0.7,$ in the previous section  (Figure \ref{fig:3D_1D_torque_summary}) we now, in Figure \ref{fig:results_1D_smoothed}, visually compare radial metrics from 2D simulations using potentials $\Phi_2, b = 0.4$; $\Phi_2, b = 0.6$; $\Phi_4, b = 0.7$, and $\Phi_4, b = 1.1$ with 3D simulations. All of these 2D potentials have similar qualitative features, however $\Phi_2, b=0.4$ and $\Phi_4, b = 0.7$ simulations, dramatically overestimate both torque excitation and gap growth. Of the simulations shown, $\Phi_2$ with $b=0.6$ approximates 3D results the best, although those simulations still shows differences from 3D in gap spacings and depth.

In Figure \ref{fig:val_over_b_second}, we show how summary metrics differ as a function of smoothing length for $\Phi_2$ and $\Phi_4$. Summary metrics, other than the migration torque, are independent of $p$ and $q$. $\Tosl$ and $x_s$ both decrease with smoothing length, whereas gap spacing increases with smoothing length. We interpolate these values against our 3D measures to identify the optimal smoothing length for each metric. These are listed in Table \ref{tab:optimal_b}. We find that optimal smoothing lengths are significantly dependent on the metric of choice. Gap spacing and $\Tosl$ prefer significantly different smoothing lengths, which is concerning given that these are the physical values that are most likely to directly impact observations \citep[$\Tosl$ determines steady state gap depths, e.g.][]{kanagawa_formation_2015}.

In Figure \ref{fig:torque_over_b_second}, we plot migration torques as a function of smoothing length and background gradient. Generally, migration torques increase with decreasing smoothing lengths, but as the gradients of the torques also change there is no single smoothing length that can match the torque structure of the 3D simulations. This confirms expectations from linear theory \citep{tanaka_three-dimensional_2024}.


\section{Discussion}
\label{sec:discussion}


\begin{figure}
    \centering
    \includegraphics[width=\linewidth]{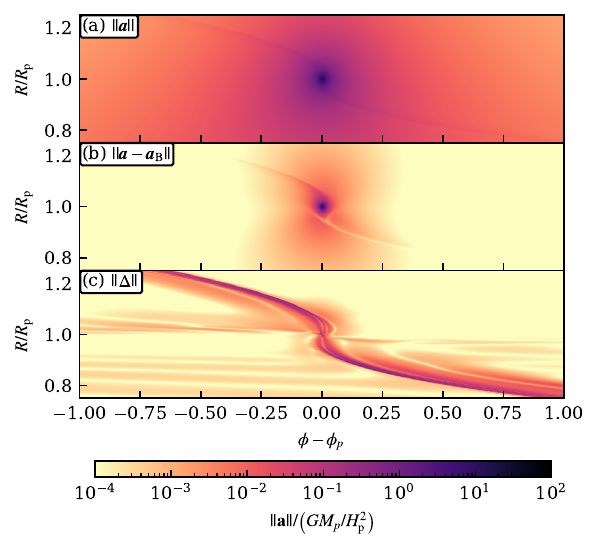}
    \caption{
    A comparison between the magnitude of 3D gravitational acceleration (a), the magnitude of the difference between actual 3D acceleration and acceleration calculated assuming
    that the vertical density profile doesn't change from the initial state, i.e. the `2D' gravitational acceleration (Equations \ref{eq:fb_r} and \ref{eq:fb_phi}) using $\Sigma$ as calculated from the 3D simulation, (b), and the remainder acceleration term (Equation \ref{eq:Delta}) which describes the acceleration of the `2D' fluid velocities caused by  inherently 3D motions (c). These have all been computed from the $p=1.5, q=0$ fiducial simulation.
    }
    \label{fig:relative_deviation}
\end{figure}

\begin{table}
\caption{Average proportional systematic errors associated with 2D simulations that use different forms of the planetary potential as compared to our 3D simulations. This is defined as $\frac{y_{\mathrm{2D}}(p, q) - y_{\mathrm{3D}}(p, q)}{y_{\mathrm{3D}}(p, q)}$ for some value $y$, which is then averaged over all values of $p$ and $q$ where 2D simulations exist. We use this definition rather than the RMS error to retain the systematic direction of the error.}
\label{tab:fit_err}
\begin{tabular}{lrrrr}
\toprule
 & $\frac{\Delta T}{T_{\mathrm{3D}}}$ & $\frac{\Delta \Tosl}{T_{\mathrm{OSL, 3D}}}$ & $\frac{\Delta (\Rog - \Rig)}{R_{\mathrm{OG1, 3D}} - R_{\mathrm{IG1, 3D}}} $ & $ \frac{\Delta x_\mathrm{s}}{x_\mathrm{s, 3D}} $ \\
\midrule
$\Phi_{\mathrm{B}}$ & -0.34 & -0.11 & 0.08 & -0.04 \\
$\Phi_{\mathrm{B}, \Hp}$ & -0.05 & -0.10 & 0.07 & -0.04 \\
$\Phi_2, b= 0.4$ & 0.22 & 0.41 & -0.04 & 0.03 \\
$\Phi_2, b=0.6$ & 0.06 & 0.05 & 0.10 & -0.04 \\
$\Phi_2, b=0.7$& -0.02 & -0.10 & 0.18 & -0.06 \\
$\Phi_4, b=0.7$ & 0.26 & 0.45 & -0.01 & 0.02 \\
$\Phi_4, b=1.1$ & -0.12 & -0.01 & 0.22 & -0.06 \\
\bottomrule
\end{tabular}
\end{table}

In the previous section we showed that there are some fundamental differences between 2D and 3D simulations of planet--disc interactions, but that 2D simulations are still able to provide good predictions for gap profiles.  From these analyses we suggest that $\Phi_{\mathrm{B}, \Hp}$ is the optimal way to treat the planetary potential in 2D simulations -- as even though $\Phi_{\mathrm{B}}$ may be more theoretically justified, $\Phi_{\mathrm{B}, \Hp}$ provides a more accurate torque approximation. We find, generally however, that 2D simulations are not capable of capturing all of the physical ingredients leading to the migration torque, and do not reproduce horseshoe vortensity striping.

To understand the origin of these differences, we should go back to Section \ref{sec:integrated_ns} and the discussion after equation (\ref{eq:NS3}), where we identified two key factors making 2D and 3D fluid equations different in isothermal discs: the ${\bf \Delta}$ term given by equation (\ref{eq:Delta}) and the deviation of the modelled planetary acceleration (e.g. $\bfa_\mathrm{B}$ given by equations (\ref{eq:fb_r}), (\ref{eq:fb_phi}) if one assumes that the vertical density distribution is unchanged) from the true vertically-integrated 3D acceleration $\bfa$, see equation (\ref{eq:acc2D}). In Figure \ref{fig:relative_deviation}, we plot in panel (b) the magnitude of the difference between $\bfa$ and $\bfa_\mathrm{B}$ obtained from either the density of a 3D simulation or the surface density of that same simulation, to be compared with $||\bfa||$ itself, shown in panel (a); both are normalized by $G\Mp/\Hp^2$, a characteristic acceleration at the distance of $\Hp$ from the planet. The deviation from $\bfa$ is most important within one $\Hp$ of the planet where the hydrostatic envelope forms driving deviations of $\rho$ from $\rho_0$ (assumed in computing $\bfa_\mathrm{B}$). Outside this region the differences are small, meaning that there $\bfa_\mathrm{B}$ approximates $\bfa$ quite well. We also show a map of $|{\bf \Delta}|$ in panel (c), which demonstrates that the remainder term is significant mainly around the planet-driven density wake (and also likely contributes to horseshoe vortensity striping). 
The non-trivial contributions provided by these two factors in different parts of the disc are what ultimately precludes the exact correspondence between the results of 2D and 3D simulations.

\postreview{It is possible to create a `more correct' form of $\bfa$ by including some of the effects of the planet on the vertical density distribution. \cite{muller_treating_2012} attempted this by estimating an adjusted scale height in the vicinity of the planet. However, while this is {\it possible}, across the majority of the disc the deviations due to ${\bf \Delta}$ are of a similar order of magnitude as the deviations due to mismatches in gravitational acceleration, and as such adding additional complexity to $\bfa$ would not necessarily make the convergence between 2 and 3D simulations significantly better.
}


\subsection{Interpretation of observations}
\label{sec:obs}


To understand how using 2D simulations may affect the  interpretation of planet--disc observations, we present the proportional systematic error for our various metrics based on comparison between 2D and 3D simulations in Table \ref{tab:fit_err}. 
While the values presented in the table may not generalise directly with changes in planetary mass and other physical parameters, we believe it can be used as an initial estimate of the systematic bias caused by 2D approximations. 
For example, simulations using $\Phi_2$ potential with $b = 0.7$, as suggested by \cite{muller_treating_2012}, would overestimate gap spacing by $18$ per cent (as shown in Figure \ref{fig:3D_1D_torque_summary}). If we attempted to connect this to planet mass using a method such as in \cite{dong_multiple_2018}, who also used $\Phi_2$ potential with $b = 0.7$, we would underestimate the planet mass by a factor of 2.2. 
Using the same methodology with $\Phi_2, b=0.6$, another common potential used by e.g  \cite{zhang_disk_2018}, \cite{ruzza_dbnets_2024}, \cite{wu_using_2023}, 
would lead to an underestimate of planet mass of order $30$ per cent\footnote{We wish to emphasise that this does not discount their important results of \cite{dong_multiple_2018}, gap spacing/width remains the likely best measure of planetary masses in continuum observations. We use their work as an example as they have provided a formulae from which we can estimate the systemic error, this error will be shared by any other method that uses gap spacing as a key observable.
}. Our analysis here is based on gas, rather than dust, dynamics and to fully verify the claim of this paragraph one would need to perform inviscid 3D gas-dust simulations. In addition this error is still smaller than those expected from observational uncertainty alone. Nonetheless, as this error is systematic rather than random, it may in future cause statistically significant biases.  For papers that use $\Phi_4, b=0.7$, \cite[e.g.][]{cimerman_planet-driven_2021, cimerman_emergence_2023, rafikov_vortex_2023, cordwell_early_2024} the numerical effects are even more significant, and $\Tosl$ in those papers would have been underestimated by a factor of two, leading to significant impacts on any estimated timescales. 

Unlike \cite{paardekooper_torque_2010, paardekooper_torque_2011} we find that $b = 0.4$ does not generally provide a good match for migration torques, except when $p=0.5$ and $q=0$, in addition it greatly overestimates the one-sided Lindblad torque, indicating that it is not capturing the overall physics of the interaction correctly. In fact, we find that no 2D potential is capable of reproducing the scalings the migration torque has with background disc gradients. While this may not be an issue for the very low $\Mp$ case where well developed models for the 3D Type 1 torque exist \citep{dangelo_three_2010, tanaka_three-dimensional_2024}, the torque on planets of moderate mass, such as $\Mp = 0.25 \Mth$, differ from linear expectation \citep{masset_on_2006} and are not as well established. \postreview{
We find that the standard linear calculation from \citet[][see their  equation (63)]{tanaka_three-dimensional_2024} does not provide a good match to our measured 3D migration torques. However, for the \emph{globally} isothermal case there is an acceptable approximation if the horseshoe -- rather than linear co-rotation -- torque \citep[equation 71 in][]{tanaka_three-dimensional_2024} is used. Nonetheless, for discs with a non-zero background gradient in temperature the quoted approximation does not work well likely due to non-linear dynamics. These differences cannot be attributed to gap opening or partial Type II migration, as when we measured the migration torques the depth of the gap was less than one percent of the background surface density.}

Based on Tables \ref{tab:fit_err} and \ref{tab:optimal_b}, we suggest that, if the Bessel-type potential is not used, that the second order smoothed potential with $b=0.64$ is the best alternative. This is very similar to the value of $b = 0.65$ already used as default by the popular code \textsc{FARGO3D} \citep{benitez-llambay_FARGO3D_2016}.  We stress that papers using potentials with non-optimal smoothing will still represent qualitatively correct physics, however the specific numerical values reported by those studies may be different from 3D reality. To study migration torques, there is no specific best fit value of $b$, as it is highly dependent on $p$ and $q$. It is not sufficient, either, to chose a value of $b$ based on the initial background indices, as in most cases this choice will lead to inaccurate estimates of $\Tosl$ and therefore gap development. As gap depth at any given time can strongly affect the total torque on a planet,  this too will lead to inaccurate overall migration.

The differences in total torque between the 2D and 3D cases may also affect long term gap structure.
While in the early stages of gap opening the total torque does not have a significant effect on gap structure, at viscous steady state the amplitude of the density pile up exterior to a planetary orbit is determined by the total torque on the planet \citep{dempsey_pileups_2020}. Therefore effects that are dependent on the scale of this pile up such as pebble isolation masses \citep{bitsch_pebble_2018}, or stalling radii for actively accreting massive planets \citep{scardoni_inward_2022}, may be different in 3D than in existing 2D analysis simply because the total torque is different.


\subsection{Implications for velocity kink observations}


\begin{figure}
    \centering
    \includegraphics[width=\linewidth]{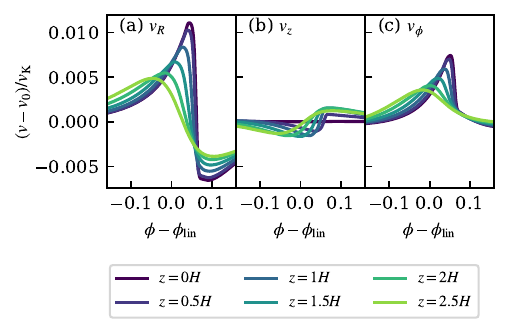}
    \caption{
    Velocity structure near the shock front at $R = 1.1\Rp$, as a function of altitude above the midplane, for our fiducial 3D simulation. 
    }
    \label{fig:vel_over_height}
\end{figure}

3D effects also impact the structure of observed velocity kinks \citep[e.g.][]{disk_dynamics_collaboration_visualizing_2020}. In Figure \ref{fig:shock_height} we showed how the density of the planetary wake changes with altitude, and now in Figure  \ref{fig:vel_over_height} we show how fluid velocity perturbations near the shock front vary with $z$. One can see that velocity perturbations in $R$ and $\phi$ show monotonic decrease in amplitude with $z$, by around a factor of 2 going from the midplane to $z=2H$, while the deviation in $v_z$ reveals a more complicated behaviour. 

There are two key effects in 3D that can change interpretation of velocity kink observations. Firstly, since the magnitude of the velocity signal is smaller one scale height above the disc (where we expect typical velocity tracers, such as the $^{12}$CO line, to be emitted) than at the midplane, interpretation of kinematic signatures using 2D simulations may underestimate the underlying planet mass. Secondly, while the magnitude of $v_z$ is smaller than that of $v_R$ or $v_{\phi}$ (at one scale height above the mid-plane its magnitude is approximately $20\% v_R$), it may still be the dominant component of the line of sight velocity deviation for very face-on systems with disc inclinations up to $\sim 10^\circ$. While some existing studies of velocity kinks use fully 3D simulations \citep[e.g][]{pinte_kinematic_2019}, others use underlying 2D hydrodynamic models \cite[e.g.][]{gardner_exoalma_2025}, and analytical studies have often focused on how planet--disc interactions affect $v_{\phi}$ only \citep[e.g.][]{wolfer_exoalma_2025}. We caution that excluding altitude dependence of velocity perturbations and the vertical velocity component may significantly misrepresent the expected signals from velocity kinks, although we leave a thorough investigation of these features to future work.


\subsection{Future Work}


A key limitation of this study is the adoption of the locally isothermal equation of state and have ignored radiative effects. The second paper in this series, \cite{ziampras_in_prep}, \postreview{addresses} this concern through a comparison of 2D and 3D simulations with radiation or beta-cooling enabled.
\postreview{
We also only tested a single planetary mass in our simulations. While we do not expect the main results to depend on mass -- we do not include any explicit dependence on the planetary mass in our smoothed potentials and our recommended potential does not require the fitting of a free parameter -- we may see more subtle effects, including different dependencies on the systematic error, at different masses. This is likely to be of highest importance for calculations of migration rates, which even at subthermal masses can deviate significantly from purely linear calculations  \citep[e.g][]{masset_on_2006}.
}

\begin{figure}
    \centering
    \includegraphics[width=\linewidth]{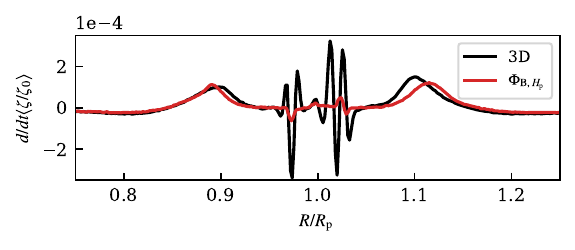}
    \caption{
    Time derivative of integrated vortensity as measured from a 3D simulation and a 2D simulation using the potential $\Phi_{\mathrm{B, \Hp}}$ between orbits 10 and 11. The additional movement of vortensity in 3D simulations may complicate the dynamical co-rotation torque. 
    The time averaged evolution of vortensity in this region is somewhat smaller than this plot suggests though there is still signifiant evolution of vortensity in this region for at least the first 100 orbits of the simulation. }
    \label{fig:vortensity_ev}
\end{figure}

In addition, we have not considered how dust dynamics may be affected by 3D effects. As dust often settles closer to the mid-plane the average gravitational force may be larger than for the gas, and it is unclear how this will change 3D circulation.
Recently, \cite{restrepo_self-gravity_2025} derived a Bessel-based formalism to represent 3D gravity in 2D simulations of self-gravitating discs without planets. Their force prescription is equivalent to, though more general than, our $\FB$ definition (equations \ref{eq:fb_phi}, \ref{eq:fb_r}). It remains to be seen if their force prescription will introduce the same types of qualitative error for dust dynamics that we saw for gas in simulations using $\FB$ will occur (see, e.g., Figure \ref{fig:results_2D}). Separately for low mass planets, non-axisymmetric dust structures can cause a reversal of migration direction \citep{benitez-llambay_torques_2018}, but this process has  only been studied in 2D simulations. This process should be re-analysed in 3D, as we found that there are significant torque differences in 2D versus 3D discs.

While we have not considered migrating planets in this work we advise that the `horseshoe vortensity striping' feature may modify the dynamical co-rotation torque (DCT) \citep[e.g.][]{paardekooper_corotation_2009}. The DCT occurs due to conservation of vortensity within the planets horseshoe region, which as the planet migrates through the disc causes a contrast between the vortensity in the horseshoe region and the background vortensity of the disc. However unlike in 2D models, in 3D vortensity is not conserved. In Figure \ref{fig:vortensity_ev}, we show the generation of additional vortensity in the horseshoe region due to horseshoe vortensity striping. We checked that this vortensity generation was not merely a transient effect, and found that while it did start to saturate over a number of libration timescales, that it is possible that it would affect longer term vortensity evolution. We leave in-depth analysis of this vortensity growth and its relationship to the DCT to future work.

Finally, we have not attempted to create semi-analytical formulae for shock formation or angular momentum deposition in 3D discs (such those derived for  2D discs using $\Phi_4, b=0.7$ in \cite{cimerman_planet-driven_2021}). This is well outside of the scope of this paper, but creating such a model would be useful both for population synthesis models or for the modelling of velocity kinks.


\section{Summary}
\label{sec:summary}


In this paper we investigated the differences between 2D and 3D representations of isothermal planet--disc interactions through a large grid of hydrodynamic simulations, including tests of several different possible representations of the effective planetary gravity in 2D. 

\begin{itemize}
    \item Reduction of 3D fluid equations to 2D is a non-trivial exercise that does not result in direct correspondence between the two limits. Some simplifying assumptions are necessary to justify a 2D approach.
    \item There is an overall similarity between gap opening in 2D and 3D discs, however in 3D shock formation is altitude dependent. In 3D discs, there is also a significant vortensity evolution in the horseshoe region. 
    \item We find that no 2D simulation using accelerations or potentials explored in this work is capable of correctly reproducing the migration torque caused by a planet across a range of gradients in background surface density and temperature. 
    \item In 2D simulations, we find that the optimal way to represent the planetary potential is with the Bessel-type constant smoothing length potential $\Phi_{\mathrm{B}, \Hp}$, Equation (\ref{eq:bessel_const_h}), as initially suggested by \cite{brown_horseshoes_2024}.
    \item For the standard second and fourth order smoothed potentials, the optimal smoothing length is dependent on the metric of choice. In order to correctly match the one-sided Lindblad torque, the optimal smoothing length for the second order potential is $b = 0.65$, and for the fourth order potential, $b = 1.09$.
    \item Existing observational analyses of gap profiles may be underestimating the masses of planets carving them by anywhere from $\sim 30$ to $220$ per cent if they rely on standard 2D simulations.
    \item Interpretation of observations of velocity kinks is likely to be strongly impacted by the 3D structure of the planet-induced perturbation. At low inclinations the vertical velocity component may be the strongest contributor to the signal. For all discs, the vertical dependence of both azimuthal and radial velocity will reduce the observed magnitude of those velocities compare to velocities derived from 2D simulations.
\end{itemize}

In addition, we developed an interoperable software package, described in Section \ref{sec:software}, which can extract a large range of important metrics from planet--disc simulations run with different codes. This software package has been made freely available on GitHub: \url{https://github.com/cordwella/disc_planet_analysis}.


\section*{Acknowledgements}

A.J.C. is funded by the Royal Society of New Zealand Te Apārangi and the Cambridge Trust through the Cambridge-Rutherford Memorial Scholarship. R.R.R. acknowledges financial support through the STFC grant ST/T00049X/1 and the IAS.
We thank Cristiano Longarini and Samuel Turner for useful discussions, Gennaro D'Angelo and Steve Lubow for their support in the use of their torque model from their 2010 paper, Ruobing Dong for clarifying the smoothed potential used in his 2018 paper, Nicolas Cimerman for an adapted version of \textsc{Athena++} which we further modified for this paper, and Hidekazu Tanaka for useful comments about migration torques.

\textit{Software:} \textsc{NumPy} \citep{harris_array_2020}, \textsc{SciPy} \citep{virtanen_scipy_2020}, \textsc{Matplotlib} \citep{thomas_a_caswell_matplotlibmatplotlib_2023}, \textsc{Athena++} \citep{stone_athena_2020} and \textsc{cephes} \citep{moshier_methods_1989}.


\section*{Data Availability}
Simulation data will be made available upon reasonable request to the corresponding author. 
A version of \textsc{ATHENA++} with the Bessel-type potentials implemented for planet disc interactions, and the analysis codes developed for this work can be found on GitHub under an open source license to assist future researchers:  \url{https://github.com/cordwella/athena-disk-planet} and \url{https://github.com/cordwella/disc_planet_analysis}.



\bibliographystyle{mnras}
\bibliography{example} 



\appendix


\section{Vertical mode decomposition for the linear problem}
\label{sec:vertical_mode}


In Section \ref{sec:integrated_ns}, we defined a 2D disc through integration over the vertical structure of a 3D disc. In this section we provide a different approach. Here we use the linearised equations of motion for a 3D disc, which we decompose into vertical modes. The 2D disc is then the part of the disc represented by the lowest order vertical mode. We follow the approach \cite{tanaka_three-dimensional_2002} although, as we have no need to solve the final linear equations, we perform only vertical mode decomposition and ignore the azimuthal decomposition into Fourier modes.

We let the background density, pressure and velocity structure of the disc be represented by $\rho_0, p_0,$ and $ \mathbf{v}_0$, where 
\begin{align}
    \rho_0(R, \phi, z) &= \frac{\Sigma_0(R)}{\sqrt{2 \uppi} H} \exp \left( -\, \frac{z^2}{2H^2} \right), \\
   \mathbf{v}_0 &= (0, R \Omega, 0) \text{, and} \\ 
    \Omega &= \Omega_K \left\{ 1 - \frac12 \left( \frac{H}{R}\right)^2 \left[\frac32 + p + \frac{q}{2} \left( \frac{z^2}{H^2} + 1 \right)\right]\right\}. 
\end{align}
We retain the locally isothermal equation of state and definition of $\Sigma_0$ from Equation (\ref{eq:background_state}).

The deviation from the background state is written as $p_1 = p - p_0$, $\eta \equiv p_1/\rho_0$, $\eta' = \eta + \Phi_p$ and $\mathbf{u} = \mathbf{v} - \mathbf{v}_0$ (n.b. in this Appendix $\mathbf{u}$ represents a different quantity than in the main text of the paper). Using these definitions the linearized Navier--Stokes equations are
\begin{align}
\label{eq:lin_R_ns}
    \frac{\upartial u_{R}}{\upartial t} + \Omega \frac{\upartial u_R}{\upartial \phi} - 2 \Omega u_{\phi} &= - \frac{\upartial \eta'}{\upartial R} - \frac{q}{R} \eta, \\ 
    \label{eq:lin_phi_ns}
    \frac{\upartial u_{\phi}}{\upartial t} + \Omega \frac{\upartial u_{\phi}}{\upartial \phi} + 2 B u_R &= - \frac{1}{R} \frac{\upartial \eta'}{\upartial \phi} \text{, and} \\ 
    \label{eq:lin_z_ns}
    \frac{\upartial u_z}{\upartial t} + \Omega \frac{\upartial u_z}{\upartial \phi} &= - \frac{\upartial \eta'}{\upartial z},
\end{align}
where $B \equiv \Omega + R/2 \frac{\upartial\Omega}{\upartial R}$ is the Oort constant. 

We then expand $\eta'$ and $\mathbf{u}$ into Hermite polynomials, where for any variable $x$,
\begin{align}
    x(R, \theta, t, z) 
    &= \sum_{n=0}^{\infty} x_n(R, \theta, t) H_n(Z) \text{ with } \\
    x_n(R, \theta, t) &\equiv \frac{1}{\sqrt{2 \uppi} n!} \int_{-\infty}^{\infty}  
    \mathrm{e}^{-Z^2/2} x(R, \theta, t, HZ) H_n(Z)\dz,
\end{align}
where $Z \equiv z/H$ and $H_n$ is the relevant Hermite polynomial defined as in \cite{tanaka_three-dimensional_2002}. 
Under these definitions, the surface density deviation, and the z-averaged velocities, are determined solely by the zeroth order vertical mode, i.e. $p_{1, 0}, u_{R, 0}$, alone. 

Writing $H \propto R^{\mu}$, with $\mu = (3 - q)/2$,
we insert the Hermite expanded velocity and pressure perturbations to Equations (\ref{eq:lin_R_ns}) - (\ref{eq:lin_z_ns}). Using the identities and recursions of the Hermite polynomials from equations 17 and 20 in \cite{tanaka_three-dimensional_2002} we find:
\begin{align}
    \frac{\upartial u_{R, n}}{\upartial t} + \Omega \frac{\upartial u_{R, n}}{\upartial \phi} - 2 \Omega u_{\phi, n} &=  - \frac{\upartial \eta'_n }{\upartial R} - \frac{q}{R} \eta_n  + \eta_n' n \frac{\mu}{R} \nonumber \\ & \, \, \,\,\,\,\, + \eta_{n + 2}' (n + 2) \frac{\mu}{R}  (n + 1), \\ 
    \frac{\upartial u_{\phi, n}}{\upartial t} + \Omega \frac{\upartial u_{\phi, n}}{\upartial \phi} + 2 B u_{R, n} &= - \frac{1}{R} \frac{\upartial \eta_{n}'}{\upartial \phi} \text{, and}  \\
    \frac{\upartial u_{z, n}}{\upartial t} + \Omega \frac{\upartial u_{z, n}}{\upartial \phi} &= \frac{\upartial \eta'_n }{\upartial z} + \frac{\eta'_{n + 1}}{H} (n + 1).
\end{align}
In this step we have chosen to ignore the small z-dependence in the Oort parameter and in $\Omega$. This means we are ignoring some of couplings between the different vertical modes of $u_R$, as this will not change the gravitational force terms we can ignore this for simplicity without modifying the results from this section.

The zeroth order, or 2D mode, of the linearized equations are:
\begin{align}
    \label{eq:lin_0_mode_R}
    \frac{\upartial u_{R, 0}}{\upartial t} + \Omega \frac{\upartial u_{R, 0}}{\upartial \phi} - 2 \Omega u_{\phi, 0} &= - \frac{\upartial \eta_{0} }{\upartial R} - \frac{q}{R} \eta_{0}  +2 \eta_2 \frac{\mu}{R} \nonumber \\
    & \, \, \, \, \, \, \, \, - \frac{\upartial \Phi_{p, 0} }{\upartial R} + 2 \Phi_{p, 2}\frac{\mu}{R}, \\
    \label{eq:lin_0_mode_phi}
    \frac{\upartial u_{\phi, 0}}{\upartial t} + \Omega \frac{\upartial u_{\phi, 0}}{\upartial \phi} + 2 B u_{R, 0} &= - \frac{1}{R} \frac{\upartial \eta_{0}}{\upartial \phi}  - \frac{1}{R} \frac{\upartial \Phi_{p, 0}}{\upartial \phi} \text{, and}\\
    \label{eq:lin_0_mode_z}
   \frac{\upartial u_{z, 0}}{\upartial t} + \Omega \frac{\upartial u_{z, 0}}{\upartial \phi} &= \frac{\eta_{1}}{H} + \frac{\Phi_{p, 1}}{H}.
\end{align}
This selection of the zeroth order mode is formally the global equivalent of the local approach in \cite{brown_horseshoes_2024}. In their local, shearing box, approach all terms proportional to $\mu$, i.e. the coupling to other vertical modes in Equations (\ref{eq:lin_0_mode_R}) and (\ref{eq:lin_0_mode_phi}), were removed.

For a disc that is symmetric around the midplane, the odd Hermite modes are uniformly zero. In the case that the initial state of the disc is symmetric and the planet's orbit is not inclined, the odd modes will be zero for all time, and therefore $u_{z, 0} = 0$ for all time. Except in the case that $\mu = 0$, which implies an unlikely temperature gradient of $p = 3$, the lowest order mode cannot be decoupled from higher order modes. It is impossible to calculate the lowest order mode accurately without at least including some of the higher order modes. Regardless, we can still calculate the gravitational force on this mode and assume that $\eta_2 = 0$ to get a closed set of equations for $u_{R, 0}$ and $u_{\phi, 0}$.

Thus, to evaluate the gravitational force on a 2D disc, or somewhat equivalently the lowest order vertical mode of a 3D disc, we need to know $\Phi_{p, 0}$, $\Phi_{p, 1}$ and $\Phi_{p, 2}$.  As the planet we consider has no inclination $\Phi_{p, 1} = 0$, and the other two are evaluated as follows:
\begin{align}
    \Phi_{p, 0} &= \frac{-1}{\sqrt{2 \uppi} H} \int_{-\infty}^{\infty} \mathrm{e}^{-z^2/2H^2} \frac{G M_p}{\sqrt{s^2 + z^2}}\dz \nonumber \\
    &= \frac{-G M_p}{\sqrt{2 \uppi} H} \mathrm{e}^{y} K_0(y),
    \label{eq:phi_p_0}
    \\
    \Phi_{p, 2} &= \frac{-1}{\sqrt{2 \uppi} 2 H} \int_{-\infty}^{\infty} \mathrm{e}^{-z^2/2H^2} \frac{G M_p}{\sqrt{s^2 + z^2}} \left( \left(\frac{z}{H} \right)^2 - 1 \right)\dz \nonumber \\
    &=- \,\frac{\Phi_{p, 0}}{2}   + \frac{GM_p}{\sqrt{2 \uppi} 4 H} s^2 \mathrm{e}^{y} \big( K_0\left ( y\right) - K_1 \left( y \right) \big).
\end{align}
Where $s$ is the polar distance as defined in Equation (\ref{eq:polar_distance}) and $y \equiv s^2/4H^2$. 
Equation (\ref{eq:phi_p_0}) is the same as the Bessel potential, $\Phi_{\mathrm{B}}$ as defined in Equation (\ref{eq:besselpotl}). The gravitational acceleration in equations (\ref{eq:lin_0_mode_R}), (\ref{eq:lin_0_mode_phi}), i.e. the terms involving $\Phi_{p, 0}$ and $\Phi_{p, 2}$, are
\begin{align}
    \bfa_\mathrm{B,lin} \cdot \hat{R} &= - \frac{\upartial \Phi_{p, 0} }{\upartial R} + 2 \Phi_{p, 2}\frac{\mu}{R} \nonumber\\
    &= 
    \frac{-GM_p}{2 H^3 \sqrt{2 \uppi} R} \mathrm{e}^{y} \bigg( \nonumber \\
    & \,\,\,\,\,\,\,\,\,\,\,\,\,\,\,\,\,\, K_0(y) \left( \mu s^2 + 6 \mu H^2 - R^2 + R R_p \cos{\left(\phi - \phi_p\right)}\right) \nonumber \\
    & \,\,\,\,\,\,\,\,\,\,\,\,\,\,\,\, - K_1(y) \left( \mu s^2 - R^2 + R R_p \cos{\left(\phi - \phi_p\right)} \right)\bigg),\\
\bfa_\mathrm{B,lin} \cdot \hat{\phi} &= 
    - \frac{1}{R} \frac{\upartial \Phi_{p, 0}}{\upartial \phi} \nonumber \\
    &= \frac{GM_p}{\sqrt{2 \uppi} H} \frac{R -  R_p \cos{(\phi - \phi_p)} }{2H^2} \mathrm{e}^y \bigg(
    \nonumber \\ 
    &\,\,\,\,\,\,\,\,\,\,\,\,\,\,\,\,\,\,\,\,\,\,\,\,\,\,\,\,\,\,\,\,\,\,\,\,\,\,\,\,\,\,\,\,\,\,\,\,\,\,\,\,\,\,\,\,\,\,\,\,\,\,\,\,
    K_0 \left( y \right) - K_1\left( y \right) \bigg).
\end{align}
As in the definition of $\FB$ (Equations \ref{eq:fb_r} and \ref{eq:fb_phi}), $\FBlin$ is a non-conservative acceleration and cannot be represented as the gradient of some potential.

In addition, we are able to use this framework to define the `2D mode' of the applied torque. The torque the planet applies to the disc is in general given by Equation (\ref{eq:3D_torque}), which we are able to expand into Hermite polynomials as follows
\begin{align}
    \frac{\upartial T}{\upartial R} &\equiv 
     - \, R\oint\int^{\infty}_{-\infty}
    \rho \frac{\upartial \Phi_{\mathrm{p}}}{\upartial \phi}\dz \dphi \\
    &=- \, R\oint\int^{\infty}_{-\infty} 
\sum_{n=0}^{\infty}\sum_{m=0}^{\infty} \bigg( \nonumber \\
&\,\,\,\,\,\,\,\,\,\,\,\,\,\,\,\,\,\,\,\,\,\,\,\,\,\,\,
    \frac{\Sigma_0}{\sqrt{2 \uppi}}\mathrm{e}^{-Z^2/2} \frac{\eta_n}{\cs^2}
    H_n(Z) \frac{\upartial \Phi_{\mathrm{p}, m}}{\upartial \phi} H_m(Z) \bigg)dZ \dphi.
\end{align}
Which using the orthonormality condition for the Hermite polynomials \citep[equation 13]{tanaka_three-dimensional_2002} reduces to 
\begin{equation}
\frac{\upartial T}{\upartial R} = 
    - \, R\Sigma_0 \oint
\sum_{n=0}^{\infty}
    \frac{\eta_n}{\cs^2} n!
    \frac{\upartial \Phi_m}{\upartial \phi} \dphi,
\end{equation}
and thus, the zeroth order or `2D mode' of the applied torque is 
\begin{align}
    \frac{\upartial T}{\upartial R}\bigg|_{2D} &\equiv - R \Sigma_0 \oint \frac{\eta_0}{\cs^2} \frac{\upartial \Phi_{\mathrm{p}, 0}}{\upartial \phi} \dphi.
\end{align}

Noting that $\Phi_{\mathrm{p}, 0} = \Phi_{\mathrm{B}}$ (Equations \ref{eq:phi_p_0} and \ref{eq:besselpotl}) and using the definition of $\eta$, the 2D torque becomes Equation (\ref{eq:torque_2d_mode}) in the main text.


\section{Numerical Convergence}
\label{sec:grid_convergence}


\begin{figure}
    \centering
    \includegraphics[width=\linewidth]{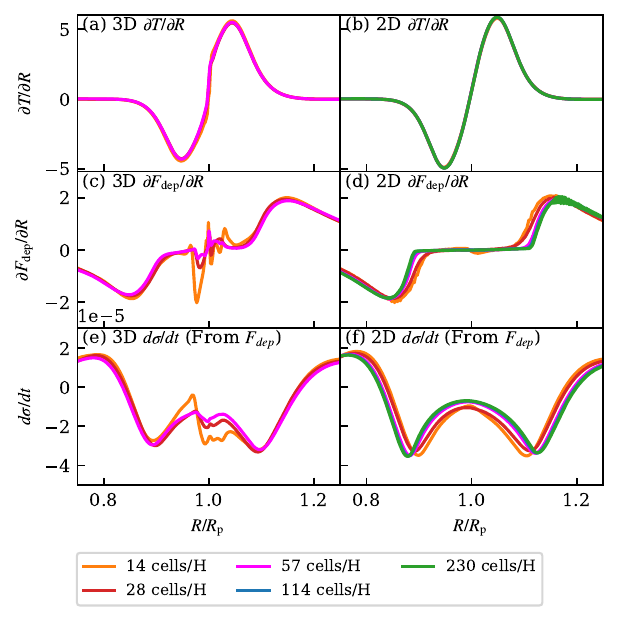}
    \caption{
    Radial metrics, see Section \ref{sec:metric_1d}, from our 2D and 3D grid convergence tests as a function of cells per scale height. The 2D simulations used the second order potential $\Phi_2, b= 0.6$. We use 28 cells per scale height as the fiducial resolution throughout this paper.
    }
    \label{fig:convergence_3d}
\end{figure}

\begin{figure}
    \centering
    \includegraphics[width=\linewidth]{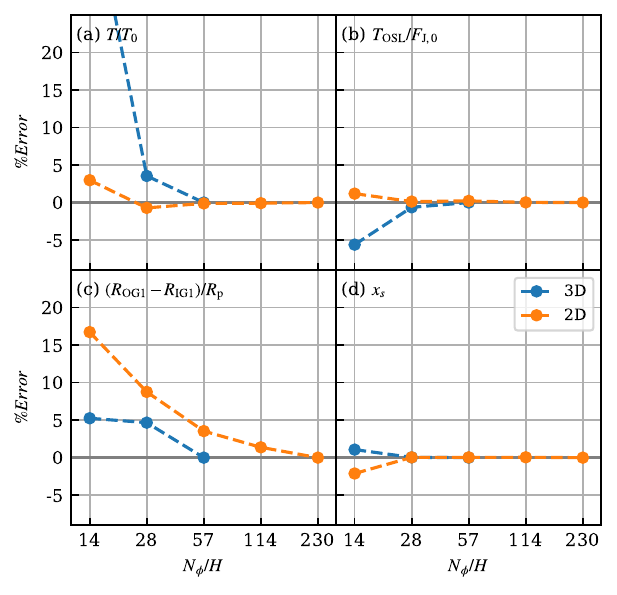}
    \caption{
    Percentage error 
    in key measures (Section \ref{sec:summary_metrics}) compared to our highest resolution simulations for 2D and 3D simulations as a function of resolution. One sided torques and horseshoe widths are converged at our fiducial resolution, however gap spacings and, for the 3D simulations, migration torques are not. 
    }
    \label{fig:convergence_summary}
\end{figure}

\begin{figure}
    \centering
    \includegraphics[width=\linewidth]{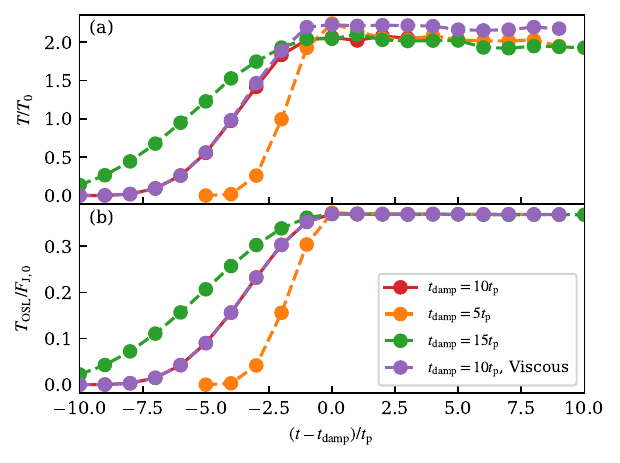}
    \caption{Tests of the temporal convergence of total and one-sided Lindblad torques for simulations with background indices $p=1.5$ and $q=0$, as a function of the ramp up time of the planetary potential $t_{\mathrm{ramp}}$}
    \label{fig:temporal_convergence_summary}
\end{figure}

To ensure that simulation dynamics are converged with respect to the numerical grid scale, we performed a set of convergence tests at increasing resolution. The simulations used surface density and temperature background indices of $p = 1.5, q=0$ for both our 2D and 3D simulations. For 2D, we used $\Phi_2, b=0.6$ as the planetary potential. We ran these tests from a lowest resolution of $(N_R, N_{\phi}) = (450,1800)$, around $14$ cells per scale height at the planet, and performed 4 additional tests, each at double the resolution of the previous simulation up to our largest simulation of $(N_R, N_{\phi}) = (7200,28800)$, i.e. 230 cells per scale height. For 3D we started at $(N_R, N_{\theta}, N_{\phi}) = (450,39,1800)$, with adjustments on the radial width in comparison to 2D to ensure they matched in cells per scale height, and performed higher resolution tests at $(N_R, N_{\theta}, N_{\phi}) = (900,78,1800)$ and $(N_R, N_{\theta}, N_{\phi}) = (1800,156,7200)$, i.e. 57 cells per scale height. We processed these simulations using the same process as in the rest of the paper, and show radial and summary metrics in Figure \ref{fig:convergence_3d} and \ref{fig:convergence_summary} respectively.

We found that the structure of $\upartial T/ \upartial R$ was converged at our lowest resolution and that the one-sided Lindblad torque and the horseshoe width were converged to a sub-percent level from $28$ cells per scale height. In 2D migration torques are well converged at the same resolution. 3D migration torques seem to be converged within $5$ per cent of their final value at this resolution, however as we did not test resolutions above $57$ cells per scale height in 3D we cannot be sure exactly when this reaches full convergence. We find that gap spacings are still not completely converged at our highest 2D resolution. We find that gap spacing increases with resolution. As resolution increases, the transition in $\upartial \Fdep/\upartial R$ from zero to non-zero values becomes sharper and occurs later - likely due to the a smaller impact of numerical viscosity on the generation of shock structure. 


Somewhat disconcertingly, the horseshoe vortensity striping feature (as discussed in Section \ref{sec:vortensity_striping}) shrinks in magnitude with increases in resolution. This can be seen near $\Rp$ in Figure \ref{fig:convergence_3d} (c) and (e). There is no significant effect of resolution on the shape and structure of this feature, but we are unsure what a completely converged version of this structure would look like.

We choose, in this paper, to use a resolution of $28$ cells per scale height to run our large grid of simulations. While this does not produce sub-percent numerical convergence for all key values, we find that this marks a good compromise between convergence and computational traceability for a large grid of simulations. This resolution is still significantly higher than many planet--disc interaction simulations.

In Figure \ref{fig:temporal_convergence_summary}, we show a plot of temporal convergence for the total and one-sided Lindblad torque for a small selection of our simulations using different ramp up times. We find that the ramp up time for the planetary potential of 10 planetary orbits is sufficient, and that the total torque and one-sided Lindblad torque are both very stable in the first 5 orbits after this point, which is when we typically measure the flow quantities. We also tested a viscous simulation, using $\nu = 2.5 \times10^{-6}$, with our standard potential ramp up time and found that the total torque was slightly more stable and slightly larger than in the inviscid case.

\section{1D Angular Momentum Evolution in a 3D Disk}
\label{sec:1d_from_3d}

\citetalias{cordwell_early_2024} derived the 1D (radial) equation for angular momentum evolution starting from 2D disc equations, which enabled them to obtain a number of important results on the early stages of gap opening. Here we extend their approach by obtaining 1D evolution equation starting from 3D fluid equations, which enables us to study gap opening based on the angular momentum deposition profiles extracted from 3D simulations. To do so, we first define the 1D averages, and the deviation from those averages, as follows:
%
\begin{align}
    \langle \Sigma \rangle (R)  &\equiv \frac{1}{2\uppi} \oint \Sigma \dphi, \\ 
    \overline{\langle v_i \rangle}(R) & \equiv \frac{1}{ 2\uppi \langle \Sigma \rangle} \oint \int \rho v_i\dz \dphi \text{, and} \\
    \delta v_i(R, \phi, z) &\equiv v_i - \langle \overline{v_i} \rangle (R),
\end{align}
for $x \in \{R, \phi, z\}$\footnote{It is also possible to derive the 1D angular momentum equations by first averaging the equations vertically, and then averaging those average velocities in azimuth, however this process leads to a more complex form of the 1D equations.}. 

3D conservation of angular momentum reads:
\begin{align}
     R \frac{\upartial}{\upartial t}(\rho R v_{\phi}) 
     &+ \frac{\upartial}{\upartial R} \left(R^2 \rho v_R v_{\phi} \right)
     +  R \frac{\upartial}{\upartial \phi} \left( \rho v_{\phi}^2 \right) + R^2 \frac{\upartial}{\upartial z} \left(\rho v_z v_{\phi} \right)
     \nonumber\\
       \label{eq:amcons}
      &= - R \frac{\upartial P}{\upartial \phi}  - R \rho \frac{\upartial \Phi}{\upartial \phi}.
\end{align}
We apply the operator $\frac{1}{2\uppi} \oint \int ....\dz \dphi$ to this equation to find:
\begin{align}
\label{eq:amf_half}
     R \frac{\upartial}{\upartial t}(R \langle \Sigma \rangle
     \langle \overline{v_{\phi}} \rangle) 
     &+ \frac{\upartial}{\upartial R} \left(R^2 
     \langle \Sigma \rangle
     \langle \overline{v_{\phi} v_{R}} \rangle \right) = \frac{1}{2\uppi} \frac{\upartial T}{\upartial R},
\end{align}
where the gravitational torque $\upartial T/\upartial R$ is defined as in Equation (\ref{eq:3D_torque}).

Using the same transformation as in Equation A9 in \citetalias{cordwell_early_2024} the second LHS term in Equation (\ref{eq:amf_half}) becomes 
\begin{align}
 R^2 \langle \Sigma \rangle
     \langle \overline{v_{\phi} v_{R}} \rangle & = 
     R^2\langle \Sigma \rangle
     \langle \overline{v_{\phi}} \rangle 
      \langle \overline{v_{R}} \rangle 
     +  
     \frac{R^2}{2\uppi} \oint\int 
     \rho \delta v_{\phi} 
      \delta v_{R}\dz \dphi \nonumber \\ 
     &=  
     \postreview{R^2} \langle \Sigma \rangle
     \langle \overline{v_{\phi}}\rangle \langle \overline{v_{R}} \rangle + \frac{1}{2 \uppi} F_{\mathrm{wave}}
     \label{eq:18}
\end{align}
where $F_{\mathrm{wave}}$ is given by Equation (\ref{eq:f_wave_3D}). Inserting the expanded term, Equation (\ref{eq:18}), into Equation (\ref{eq:amf_half}) we retrieve the standard 1D angular momentum equation:
\begin{align}
    \label{eq:base_amf}
    R\frac{\upartial}{\upartial t}(\langle\Sigma\rangle R \langle \overline{v_{\phi}} \rangle) + \frac{\upartial}{\upartial R}(R^2 \langle\Sigma\rangle  \langle \overline{v_{R}} \rangle \langle \overline{v_{\phi}} \rangle ) &= \frac{1}{2 \uppi}\frac{\upartial \Fdep}{\upartial R}  ,
\end{align}
with $\upartial \Fdep/\upartial R = \upartial T/\upartial R - \upartial F_{\mathrm{wave}}/\upartial R$ as in Equation (\ref{eq:f_dep_def}). The 1D equation of mass conservation is the same as in 2D case of \citetalias{cordwell_early_2024}. Using this equation, the theoretical surface density evolution formulae from \citetalias{cordwell_early_2024} (see their Appendix C), immediately generalize to 3D.


\section{Early stages of gap opening for different background gradients}
\label{sec:slope_on_gap}


\begin{figure}
    \centering
    \includegraphics[width=\linewidth]{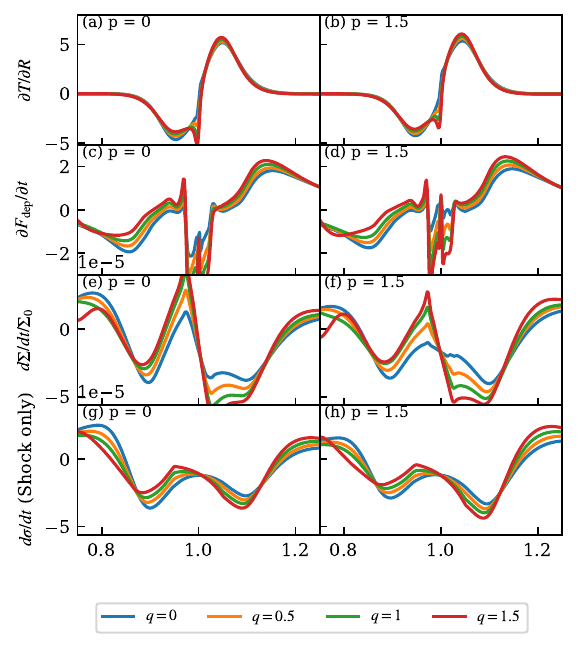}
    \caption{The same radial metrics as in Figure \protect \ref{fig:3D_1D_torque_summary}, but only showing 3D simulation results and for a larger selection of background temperature gradients. The left-hand column shows simulations with a surface density background index of $p = 0$ and the right-hand column shows those with $p = 1.5$.}
    \label{fig:gap_structure_background_density}
\end{figure}

As we have simulated a large grid of 3D planet--disc interaction problems, we are able to use this serendipitously to investigate gap formation across a wide range of background surface densities, $p$, and temperatures, $q$, thus supplementing the results of \citetalias{cordwell_early_2024}. In Figure \ref{fig:gap_structure_background_density}, we show radial metrics for eight of our 3D simulations.
We find that neither the spacing between the two gaps nor $\Tosl$ differ significantly with $p$ or $q$, although as $q$ increases the location of each gap shifts to inwards slightly. This may due to more torque being excited in the outer region than the inner as $q$ increases.

\citetalias{cordwell_early_2024} (Appendix C) showed that increasing $q$, while keeping the same angular momentum deposition profile, lead to slightly shallower inner gaps and slightly deeper outer gaps. However, the effects they measured were very small, almost unnoticeable for $h_p = 0.05$. Therefore the majority of the changes we see in gap structure must be driven by changes in $\upartial T/\upartial R$ and $\upartial \Fdep/\upartial R$, rather than through changes in how discs at different temperatures respond to angular momentum deposition.

The spikes in $\upartial T/ \upartial R$ near $\Rp$ for $p \neq 0$ seem to be related to the horseshoe vortensity striping as discussed in Section \ref{sec:results_3D}, as these were also seen in the low viscosity simulations of \cite{dangelo_three_2010}, though not in their higher viscosity simulations.


\bsp	
\label{lastpage}
\end{document}